\documentclass[preprint]{revtex4}
\usepackage{amsmath}
\usepackage{amssymb}
\usepackage{amsfonts}
\usepackage{bm}
\usepackage{graphicx}
\usepackage{color}
\newcommand{\OP}[1]{\hat{\boldsymbol{#1}}}
\newcommand{\Sch}{Schr\"odinger }
\newcommand{\BS}[1]{\boldsymbol{#1}}

\begin{document}
\title{Quantum Mechanical Results Of The Matrix Elements Of The Boltzmann
Operator Obtained From Series Representations}
\author{Mahir E. Ocak}
\email{meocak@alumni.uchicago.edu}
\affiliation{Ya\c{s}amkent Mahallesi, Yonca Sitesi 13/B Daire No:5,
	 \c{C}ayyolu, Ankara, Turkey}
\date{\today}
\begin{abstract}

Recently developed series representations of the 
Boltzmann operator are used to obtain Quantum Mechanical results
for the matrix elements,
$\langle x\vert \exp(-\beta {\hat H})\vert x'\rangle$, of the imaginary time propagator.
The calculations are done for two different potential surfaces:
one of them is an Eckart Barrier and the other one is a double well
potential surface. Numerical convergence of the series are
investigated. Although the zeroth order term is sufficient at high
temperatures, it does not lead to the correct saddle point structure
at low temperatures where the tunneling is important. Nevertheless 
the series converges rapidly even at low temperatures. 
Some of the double well calculations are also
done with the bare potential (without Gaussian averaging).
 Some equations of motion related with bare potentials
are also derived. 
 The use of the bare potential results in faster integrations of
equations of motion. Although, it causes lower accuracy in the zeroth 
order approximation, the series show similar convergence 
properties both for Gaussian averaged calculations and the bare potential
calculations. However, the series may not converge for 
bare potential calculations at low temperatures because of the low accuracy
of zeroth order approximation. 
 Interestingly, it is found that the number of
saddle points of $\langle x\vert \exp(-\beta {\hat H})\vert x'\rangle$
increases as the temperature is lowered. An explanation 
of observed structures at low temperatures 
remains as a challenge. Besides, it has implications for
the quantum instanton theory of reaction rates at very low
temperatures.
\end{abstract}

\maketitle

\section{\label{sec:intro}Introduction}
Semiclassical theories are useful for the treatments of large dimensional
systems because of their favorable scaling with the dimension of the system.
 They
allow one to study systems that are very hard to
treat with quantum mechanics due to exponential scaling of the sizes of bases
with increasing number of dimensions in quantum mechanics.
 Besides, they offer advantages
over classical methods by allowing one to include quantum effects such as
interference and tunneling which cannot be described with classical methods.

Development of semiclassical propagators for studying dynamics of atomic
and molecular systems dates back to pioneering works of Van Vleck \cite{Vleck28}
and Gutzwiller \cite{Gutzwiller}. Van Vleck Propagator is obtained as a
semiclassical approximation to path integral and it is exact for
quadratic Hamiltonians. However, it has some drawbacks due to some
difficulties related with  its
numerical implementation. Firstly, dynamics is formulated as a
double ended boundary
value problem, so that it is necessary to do nonlinear searches to find
classical trajectories that satisfy the boundary conditions of the problem.
Secondly, Van Vleck propagator has a prefactor that has singularities.
These problems with the Van Vleck propagator is overcome by the development
of initial value representation (IVR) methods. In the IVR,
integration variable related with the end point of a coordinate is
transformed to an initial momentum so that
the dynamics problem is posed as an initial value problem in which the initial
conditions of the problem are specified by the initial phase space points. Thus,
it is not necessary to make nonlinear searches for finding classical paths
connecting the end points. Besides, the IVR based propagators do not include
prefactors that have singularities. In addition to its numerical advantages
IVR also offers a more intuitive physical picture for dynamics.

History of IVR dates back to its use by Miller \cite{Miller70} and Markus
\cite{Marcus71} in studies of classical S-matrix calculations
for collisions. Modern
semiclassical propagators are based on the idea of using Gaussian wave packets
suggested by Heller \cite{Heller75,Heller81} which is later refined
by Herman and Kluk \cite{Herman84}. Since the work of Heller,
many IVR based propagators are developed for real time dynamics. Several
reviews about different aspects of the subject can be found in the literature
\cite{Herman94,Grossman96,Miller97,Miller98,Tannor2000,Baranger2001,Miller2001,Thoss2004,Kay2005}.

The success of the IVR based methods in real time dynamics also motivated the
development of semiclassical methods for imaginary time dynamics.
Several methods has been proposed for a semiclassical approximation of
the imaginary time propagator
\cite{Metiu85,Makri2002,Frantsuzov2003,Frantsuzov2004}.
Recently, Frantsuzov {\it et. al.} 
developed semiclassical approximation to imaginary time propagator 
\cite{Frantsuzov2003,Frantsuzov2004}, 
named Time Evolving Gaussian Approximation (TEGA), which is based on an earlier
method suggested by Hellsing {\it et. al.} \cite{Metiu85}.
Another approximation to imaginary time propagator is suggested by 
Pollak and Martin-Fierro \cite{Pollak2007}.
 In this method, named PSTEGA (Phase Space Time
Evolving Gaussian Approximation), coherent states are used.
Since the Gaussians that are used in the TEGA method can be considered as
coherent states with zero momentum, PSTEGA method can be considered as a
generalization of the TEGA method.  Although the 
propagator involves phase space integration instead of configuration space
integration, it is possible to integrate equations of 
motion in an efficient way. As shown in the appendix, the momentum degrees of 
freedom can be integrated implicitly so that matrix elements of the equilibrium
density matrix can be evaluated with an expansion in configuration space
as in the TEGA method. Besides, the PSTEGA method provides a new way of 
evaluating time correlation functions.

Although the IVR based
 semiclassical methods has been used successfully in many systems,
the approximations involved in these calculations remained uncontrolled
such that there was no way to estimate the errors in these calculations. 
This problem has been overcome by the development of correction operator
formalism by Pollak and co-workers 
\cite{Ankerhold2002,Pollak2003,Zhang2003-2,Zhang2003,Zhang2004,Saltzer2005}.
They have shown that the exact quantum mechanical 
propagator can be expanded in a series
in which the zeroth order term is an approximation to the exact propagator. 
Then, the higher order terms are obtained from the zeroth order term
in a recursive manner by using the correction operator. Therefore,  
it is possible obtain quantum mechanical results, at least in principle,
 starting with
a semiclassical calculation. 

In this paper, some equations of motion related with the PSTEGA calculations, 
in which bare potentials are used, are derived. In addition to that 
an efficient way of solving the PSTEGA equations of motion is shown.
Then the TEGA and the PSTEGA methods are applied to two different systems
one with an Eckart Barrier and another with a double well potential surface.
Numerical convergence properties of the series are investigated and
the accuracy of the results is checked with quantum mechanical 
calculations.  In the calculations both the Gaussian averaging and the 
bare potential are used. The results show that the use of bare
potential leads to lower accuracy in the zeroth order approximation.
Nevertheless, calculations show similar convergence properties with the Gaussian
averaged potential calculations. The results show interesting 
structures at low temperatures.

In section \ref{sec:theo}, theory of the TEGA and PSTEGA methods are
reviewed. Then, in section \ref{sec:calc} calculations and results 
are presented. The paper ends with discussions and 
conclusions in section \ref{sec:conc}.

\section{\label{sec:theo}Theory}
The theory of the TEGA method as a semiclassical approximation is
first  developed by 
Frantsuzov and Mandelshtam \cite{Frantsuzov2003,Frantsuzov2004}
based on an earlier method suggested by Hellsing {\it et.al.} \cite{Metiu85}.
 Later, Shao and Pollak applied the Correction
Operator formalism to the TEGA method and showed how it can be
used to obtain quantum mechanical results for the matrix elements
of the Boltzmann operator\cite{Shao2006}.
The theory of the PSTEGA method 
is developed by Pollak and Martin-Fierro \cite{Pollak2007}.
In the following subsections, first the theory of the correction operator
formalism and the series expansion of the thermal propagator will be given
in a general manner. Then, the details of TEGA and PSTEGA methods will be
given. Equations of motion for the PSTEGA calculation in which the bare 
potential is used will be given for the first time. 

\subsection{\label{ssec:pre}Preliminaries}
Consider an $N$ dimensional system with the Hamiltonian 
\begin{equation}
\hat{H} = \frac{\OP{p}^2}{2} + V(\OP{q}),
\end{equation}
where $\OP{q}$ and $\OP{p}$ are $N$ dimensional 
vectors of mass weighted coordinate and momenta respectively satisfying
the usual commutation relation $[\hat{q_i},\hat{p}_j]= i \hbar \delta_{ij}$, 
and $V(\OP{q})$ is the potential surface of the system. 

The thermal propagator $\hat{K}(\tau) \equiv \exp (-\tau \hat{H})$ is the
solution of the imaginary time  \Sch equation (or the Bloch equation),
\begin{equation}
\left( - \frac{\partial}{\partial \tau} - \hat{H} \right) \hat{K}(\tau) = 0,
\end{equation}
at imaginary time $\tau$ with the initial condition $\hat{K}(0)=I$,
 where $I$ is the $N$ dimensional
identity matrix. 

If there exist a good approximation $\hat{K}_0(\tau)$ for the exact propagator 
$\hat{K}(\tau)$, then the correction operator $\hat{C}(\tau)$ 
can be defined as follows:
\begin{equation}
\hat{C}(\tau)= \left( - \frac{\partial}{\partial \tau} - \hat{H} \right)
 \hat{K}_0(\tau).
\end{equation}
The differential equation above can be inverted to an integral equation 
by realizing that the exact propagator is the solution of the homogeneous
equation (Bloch equation). The formal solution is given by,
\begin{equation}
\hat{K}(\tau) = \hat{K}_0(\tau) + \int_0^{\tau} \mathrm{d}\tau' 
	\hat{K}(\tau-\tau')\hat{C}(\tau'). \label{eq:forsol}
\end{equation}

\subsection{Series Representation Of The Thermal Propagator}
Although equation (\ref{eq:forsol}) provides a formal solution, it is not very
useful in that form since the 
exact propagator appears on both sides of the equation. 
On the other hand, if $\hat{K}_0(\tau)$ is a good 
approximation to the exact propagator, $\hat{K}(\tau)$, then it makes sense to expand the exact solution in a series
where $\hat{K}_0(\tau)$ is the leading order term of the series such that
\begin{equation}
	\hat{K}(\tau) = \sum_{i=0}^{\infty} \hat{K}_i(\tau). \label{eq:kexp}
\end{equation} 
By plugging the expansion above to equation (\ref{eq:forsol}), and 
assuming that $\hat{K}_j \sim \hat{C}^j$, the following recursion relation
is obtained for the higher order terms
by equating the terms that are of the order of the same power of the 
correction operator: 
\begin{equation}
	\hat{K}_{i+1}(\tau)= \int_0^{\tau} \mathrm{d}\tau' 
		\hat{K}_i(\tau -\tau') \hat{C}(\tau'), \; \; i\geq 0.
	\label{eq:kiexp}
\end{equation}
Thus, given an approximation, the 
exact thermal propagator can be obtained from that approximation 
recursively, by using a series in which the zeroth order term is the 
approximation. 

\subsection{Symmetric Form Of The Series Representation}

The exact thermal propagator is Hermitian: 
$\hat{K}(\tau)^{\dagger} = \hat{K}(\tau)$. However, both the TEGA and the PSTEGA
approximations do not provide a hermitian representation for $\hat{K}_0(\tau)$. 
Therefore, the series expansion that contains the TEGA or the PSTEGA
approximations as the zeroth order term cannot give a Hermitian representation
of the exact propagator. This problem can be remedied as follows.  
In order to make every single term in the series expansion Hermitian,
the equation below can be used:
\begin{equation}
	\hat{K}(\tau) = \hat{K}(\tau/2) \hat{K}^{\dagger}(\tau/2). 
		\label{eq:herform}
\end{equation}
Since $(\hat{K}(\tau/2)\hat{K}(\tau/2)^{\dagger})^{\dagger}= 
\hat{K}(\tau) \hat{K}(\tau/2)^{\dagger}$, 
the use of this identity guarantees generation of a Hermitian representation of
$\hat{K}(\tau)$ from any representation of $\hat{K}(\tau/2)$ regardless of
 whether that representation 
is Hermitian or not.
Thus, by expanding all of the terms in equation (\ref{eq:herform}) in a series 
as in equation (\ref{eq:kexp}), the following series expansion is obtained 
for $\hat{K}(\tau)$ in terms of the terms in the series expansion of 
$\hat{K}(\tau/2)$
\begin{equation}
	\hat{K}(\tau)=\sum_{j} \hat{K}^{(j)}(\tau), 
\end{equation}
where 
\begin{equation}
	\hat{K}^{(j)}(\tau) = \sum_{i=0}^{i=j} \hat{K}_i (\tau/2) 
		\hat{K}_{j-i}^{\dagger}(\tau/2). \label{eq:symexp}
\end{equation}
It is clear that a representation of the term on the left hand side
will be Hermitian regardless of
whether the representations of the terms on the right hand
 side  are Hermitian or not.

\subsection{Definitions Of Averaged Quantities}
Frantsuzov and Mandelshtam derived the equations of motion for the TEGA method
variationally \cite{Frantsuzov2004}. This method leads to the result that
the potential and its derivatives should be Gaussian averaged. 
In the rest of the paper, the following notation is used for denoting the Gaussian
averaging of a quantity $h(\BS{q})$:
\begin{eqnarray}
\langle h(\BS{q})\rangle & = & \left(\frac{1}{\pi}\right)^{N/2}
    \frac{1}{\sqrt{\vert \mathrm{det}(\BS{G}(\tau)) \vert}}
\int_{-\infty}^{\infty} \mathrm{d}\BS{x} \nonumber \\
& & \times \exp(-(\BS{x}-\BS{q}(\tau))^T
        \BS{G}(\tau)^{-1}(\BS{x}-\BS{q}(\tau))) h(\BS{x}),
\label{eq:defave}
\end{eqnarray}
where $\BS{q}(\tau)$ is an $N$ dimensional vector defining the 
center of the Gaussian and $\BS{G}(\tau)$ is an $N \times N$ 
dimensional positive definite
matrix defining the width of the Gaussian.

Shao and Pollak has
shown that the equations of motion for the TEGA method can be derived by
requiring that the method is exact for harmonic potentials \cite{Shao2006},
and the same idea is also used in the development of the PSTEGA method 
\cite{Pollak2007}.
 By using this idea of deriving equations of motion,
Shao and Pollak suggested that the TEGA method can be generalized for
an arbitrary averaging function. 
Given a reasonable zeroth order approximation, the series representation
will converge to the correct result. Consequently, Shao and Pollak suggested
that the calculations can be done with the bare potential (without 
Gaussian averaging).
Provided that this still gives a reasonable zeroth order approximation, one can
obtain the exact propagator from that approximation via correction operator
formalism. The use of the bare potential leads to faster integrations of 
equations of motion since the Gaussian averagings are not done.  

\subsection{\label{ssec:emtega}Equations Of Motion For The TEGA}
Matrix elements of the approximate solution of the Bloch equation in the
TEGA is given by \cite{Frantsuzov2003,Frantsuzov2004},
\begin{eqnarray}
\langle \BS{x} \vert \hat{K}_0(\tau)\vert \BS{q}_0\rangle & = & 
\left( \frac{1}{2 \pi}\right)^{N/2} \frac{1}{\vert \mathrm{det}(\BS{G}(\tau))\vert^{1/2}} \nonumber \\ & & 
 \times \exp \left(-\frac{1}{2}((\BS{x}-\BS{q}(\tau))^T \BS{G}(\tau)^{-1} 
	(\BS{x}-\BS{q}(\tau))) + \gamma (\tau)\right). \label{eq:tega}
\end{eqnarray}
In the equation above, the $N$ dimensional vector $\BS{q}=\BS{q}(\tau)$ 
defines the center of the Gaussian, the $N \times N$ dimensional 
positive definite matrix
$\BS{G}=\BS{G}(\tau)$ defines the width of the Gaussian, and the parameter
$\gamma=\gamma(\tau)$ is a real scale factor. 

In order for this representation of the thermal propagator to satisfy the
initial condition that $\hat{K}(0)=I$, the following conditions should
be imposed for small $\tau$:
\begin{equation}
\BS{q}(\tau \simeq 0) = \BS{q}_0, \; \BS{G}(\tau \simeq 0) = \hbar^2 \tau I,
	\; \gamma(\tau \simeq 0) = -\tau V(\BS{q}_0). \label{eq:tegainicond}
\end{equation} 

If Gaussian averaging is used, the equations of motion for the three variables
are \cite{Frantsuzov2004,Predescu2005}:
\begin{eqnarray}
\frac{\mathrm{d}}{\mathrm{d}\tau} \BS{G}(\tau) & = & -\BS{G}(\tau) 
\langle \nabla \nabla^T V(\BS{q}(\tau))\rangle \BS{G}(\tau) +\hbar^2 I, 
\label{eq:tegaeom1} \\
\frac{\mathrm{d}}{\mathrm{d}\tau}\BS{q}(\tau) & = & -\BS{G}(\tau) 
	\langle \nabla V(\BS{q}(\tau))\rangle, \label{eq:tegaeom2} \\
\frac{\mathrm{d}}{\mathrm{d}\tau} \gamma(\tau)& = & 
	\frac{-1}{4}\mathrm{Tr}[\langle \nabla \nabla^T 
	V(\BS{q}(\tau))\rangle\BS{G}(\tau)]- \langle V(\BS{q}(\tau))\rangle.
\label{eq:tegaeom3}
\end{eqnarray}
The use of the bare potential leads to the same equations of motion except
that the potential and its derivatives are not Gaussian averaged.
 However, the  coefficient 
$-1/4$
in the equation of motion of the variable $\gamma(\tau)$ becomes
 $-1/2$ \cite{Shao2006}. 

 When equation (\ref{eq:symexp}) is used
the matrix elements of the each term in the series expansion of the 
Boltzmann operator can be calculated as follows:
\begin{equation}
\langle \BS{x} \vert \hat{K}^{(j)}(\tau)\vert \BS{x}' \rangle = \sum_{i=0}^{i=j}
    \int d\BS{y} \langle \BS{x} \vert \hat{K}_i(\tau/2)\vert \BS{y} \rangle
	\langle \BS{y} \vert \hat{K}_{j-i}^{\dagger}(\tau/2)\vert \BS{x}' \rangle. 
\label{eq:symmattr}
\end{equation}

\subsection{\label{ssec:empstega}Equations Of Motion For The PSTEGA}
In the PSTEGA method \cite{Pollak2007},
 the thermal propagator is represented in a 
coherent state basis whose coordinate state representation is given by
\begin{eqnarray}
\langle \BS{x} \vert g(\BS{p},\BS{q},\tau)\rangle & = & \left( 
	\frac{1}{\mathrm{det}(\BS{G}(\tau)) \pi^N}\right)^{1/4} \nonumber \\
& &	\exp \left( -\frac{1}{2} (\BS{x}-\BS{q}(\tau))^T \BS{G}^{-1}(\tau)
	(\BS{x}-\BS{q}(\tau)) + \frac{i}{\hbar} 
	\BS{p}^T(\tau)(\BS{x}-\BS{q}(\tau))\right).
\end{eqnarray}
In the equation above, $N$ dimensional vectors 
$\BS{q}=\BS{q}(\tau)$ and $\BS{p}=\BS{p}(\tau)$ are the position and momentum
vectors respectively and the $N \times N$ dimensional positive definite matrix
 $\BS{G}(\tau)$ is the width matrix.
The matrix elements of the imaginary time propagator can be expanded
in the coherent state basis as 
\begin{equation}
\langle \BS{x} \vert \exp (-\tau \hat{H})\vert \BS{x}'\rangle =
\int \frac{\mathrm{d}\BS{p}\mathrm{d}\BS{q}}{2 \pi} 
\langle x \vert \exp (-\tau \hat{H})\vert g(\BS{p},\BS{q},0)\rangle
\langle g(\BS{p},\BS{q},0) \vert \BS{x}'\rangle,
\end{equation}
and the mixed matrix elements of the thermal
propagator are approximated as 
\begin{eqnarray}
\langle \BS{x} \vert \exp(-\tau \hat{H}) \vert g(\BS{p},\BS{q},0)\rangle & 
\simeq & \langle \BS{x} \vert \hat{K}_0(\tau)\vert g(\BS{p},\BS{q},0) \rangle 
\nonumber \\  &  
\equiv & 
f(\BS{p},\BS{q},\tau) \langle \BS{x} \vert g(\BS{p},\BS{q},\tau)\rangle .
\end{eqnarray} 

If the potential is Gaussian averaged, equations for the variables,
 $\BS{q}$, $\BS{p}$, $\BS{G}$, and $f(\tau)$ are
given by \cite{Pollak2007}
\begin{eqnarray}
	\frac{\partial \BS{q}(\tau)}{\partial \tau} & = & 
 -\BS{G}(\tau) \langle \nabla V(\BS{q}(\tau))\rangle, \; \;\BS{q}(0)=\BS{q}_0,\\
	\frac{\partial \BS{p}(\tau)}{\partial \tau} & = & 
		- \hbar^2 \BS{G}(\tau)^{-1} \BS{p}(\tau), \; \; 
	\BS{p}(0)=\BS{p}_0,	\label{eq:peom} \\ 
	\frac{\partial \BS{G}(\tau)}{\partial \tau} & = & 
-\BS{G}(\tau) \langle \nabla \nabla^T V(\BS{q}(\tau))\rangle \BS{G}(\tau)
	+ \hbar^2 I, \\
f(\tau) & = & \exp \left( - \int_0^{\tau} \mathrm{d}\tau' \left[ 
	\frac{1}{2} \BS{p}^T(\tau') \BS{p}(\tau')
	+ \langle V(\BS{q}(\tau'))\rangle \right. \right. \nonumber \\
 & & \left. \left. + \frac{\hbar^2}{4} \mathrm{Tr}[\BS{G}(\tau')^{-1}] 
	- \frac{i}{\hbar} \BS{p}^T(\tau'). 
	\frac{\partial \BS{q}(\tau')}{\partial \tau'}\right]\right).
\end{eqnarray}
As the way Pollak and Martin-Fierro have done \cite{Pollak2007}, the equations
of motion for the bare potential calculations can be derived by requiring
that the method is exact for quadratic potentials. 
In this case $f(\tau)$ is 
given by 
\begin{eqnarray}
f(\tau) & = & \exp \left( - \int_0^{\tau} \mathrm{d}\tau' \left[ 
	\frac{1}{2} \BS{p}^T(\tau') \BS{p}(\tau') 
	+ V(\BS{q}(\tau')) \right. \right. \nonumber \\
 & & \left. \left. + \frac{\hbar^2}{4} \mathrm{Tr}[\BS{G}(\tau')^{-1}] 
	- \frac{i}{\hbar} \BS{p}^T(\tau'). 
	\frac{\partial \BS{q}(\tau')}{\partial \tau'} \right. \right. 
	\nonumber \\ 
	& &+ \left. \left. \frac{1}{4} \mathrm{Tr} [ \nabla \nabla^T V(\BS{q}(\tau')) 
		\BS{G}(\tau')]\right]\right).
\end{eqnarray}
Other equations will be the same except that the potential and its 
derivatives are not Gaussian 
averaged. In the PSTEGA approximation, initial width of the Gaussians
is arbitrary.

In calculations, it is not necessary to make a propagation in phase space,
since the equation of motion for $\BS{p}(\tau)$, equation (\ref{eq:peom}),
 can be integrated 
implicitly. For a detailed explanation of how to integrate equations of 
motion in an efficient way see the appendix. 

\subsection{\label{ssec:corrop}Matrix Elements Of The Correction Operator}
Matrix elements of the correction operator are given by 
\cite{Shao2006,Pollak2007}
\begin{equation}
\langle \BS{x} \vert \hat{C}(\tau)\vert \BS{q}(\tau) \rangle = 
	- \langle V_{\mathrm{anh}}(\BS{x},\BS{q},\tau) \rangle
	\langle \BS{x} \vert K_0(\tau) \vert \BS{q}(\tau) \rangle ,
\label{eq:cop}
\end{equation}
where $\langle V_{\mathrm{anh}}(\BS{x},\BS{q},\tau)\rangle$
 is the anharmonic remainder of the 
potential when it is expanded around $\BS{q}(\tau)$. If Gaussian averaging is
used, expansion of the potential surface is given by \cite{Shao2006}
\begin{eqnarray}
	V(\BS{x}) &  \equiv & \langle V(\BS{q}(\tau))\rangle
	+\frac{1}{2} ( \langle \nabla^T V(\BS{q}(\tau)) \rangle 
		(\BS{x}-\BS{q}(\tau)) \nonumber \\
	& &  + (\BS{x}-\BS{q}(\tau))^T 
		\langle \nabla V(\BS{q}(\tau))\rangle) \nonumber \\
	& & + \frac{1}{2}(\BS{x}-\BS{q}(\tau))^T 
		\langle \nabla \nabla^T V(\BS{q}(\tau))\rangle
	 	(\BS{x}-\BS{q}(\tau)) \nonumber \\
	& & - \frac{1}{4}[\langle \nabla \nabla^T V(\BS{q}(\tau))\rangle 
			\BS{G}(\tau)] \nonumber \\
	& & + \langle V_{\mathrm{anh}}(\BS{x},\BS{q},\tau) \rangle. 
\end{eqnarray}
If the bare potential is used in the calculations; then, the
 expansion of the potential surface
is given by
\begin{eqnarray}
	V(\BS{x}) &  \equiv & V(\BS{q}(\tau))
	+\frac{1}{2} ( \nabla^T V(\BS{q}(\tau))  
		(\BS{x}-\BS{q}(\tau)) \nonumber \\ & & 
		+ (\BS{x}-\BS{q}(\tau))^T 
		\nabla V(\BS{q}(\tau))) \nonumber \\
	& & + \frac{1}{2}(\BS{x}-\BS{q}(\tau))^T 
		\nabla \nabla^T V(\BS{q}(\tau))
	 	(\BS{x}-\BS{q}(\tau)) \nonumber \\
	& &  + V_{\mathrm{anh}}(\BS{x},\BS{q},\tau). 
\end{eqnarray}
In this case, matrix elements of the correction operator are given by
\begin{equation}
\langle \BS{x} \vert \hat{C}(\tau)\vert \BS{q}(\tau) \rangle = 
	- V_{\mathrm{anh}}(\BS{x},\BS{q},\tau)
	\langle \BS{x} \vert K_0(\tau) \vert \BS{q}(\tau) \rangle.
\end{equation}

\subsection{\label{ssec:qmcal}Quantum Mechanical Calculations}
In order to make a comparison of TEGA and PSTEGA calculations with a direct
quantum mechanical calculation, quantum mechanical calculations are also 
performed. In order to calculate the density matrix elements, first
the Hamiltonian matrix is diagonalized;
then, the thermal propagator is expanded in the basis of the eigenstates
of the Hamiltonian.
If $\phi_n(x)$ are the eigenstates of the Hamiltonian which are obtained
as a result of diagonalization calculation; then, the matrix elements of 
the thermal propagator can be evaluated as follows:
\begin{eqnarray}
\langle x' \vert \exp(-\beta \hat{H})\vert x \rangle & = & \sum_n 
	\langle x' \vert \exp(-\beta \hat{H}) \vert \phi_n \rangle 
	\langle \phi_n \vert x \rangle \nonumber \\
& = & \sum_n \phi_n(x') \exp(-\beta E_n) \phi_n^*(x).
\label{eq:qdenmat}
\end{eqnarray}
This calculation, in the case of Eckart Barrier, duplicates the calculations
of Miller {\it et. al.} \cite{miller2003}.
\section{\label{sec:calc}Calculations and Results}
The calculations are done for two different potential surfaces. One of them is
an Eckart Barrier and the other one is a double well potential surface.
The details of calculations and their results are given in the following 
subsections. 
\subsection{Eckart Barrier}
The configuration matrix elements of the Eckart potential were
previously studied by Miller {\it et. al.} \cite{miller2003}. 
The asymmetric Eckart
barrier potential has the form
\begin{equation}
V(x)= \frac{V_0 (1-\alpha)}{1+\exp(-2ax)} +
    \frac{V_0 (1+\sqrt{\alpha})^2}{4\cosh^2(ax)} \label{eq:asmeck}
\end{equation}
where $\alpha$ is the asymmetry parameter. The potential surface is
symmetric when $\alpha=1$. In the present study, the same
parameters that were used by Miller {\it et. al.} \cite{miller2003}, that is $V_0=0.016$a.u.,
$a=1.3624$a.u. and $m=1061$a.u. are used. Some computations
for an asymmetric barrier with $\alpha=1.25$ are also performed.

Since this is a one dimensional system, it is convenient
to work with matrices of the zeroth order Boltzmann operator, equation
(\ref{eq:tega}), and the correction operator, equation (\ref{eq:cop}), in the
configuration space. Then, the final results are obtained by matrix
multiplications and time integrations. The latter were all performed
using the third order Simpson integrator \cite{NRC}. An
evenly spaced grid is taken for the coordinate $y$ in a finite
symmetric range, and a Gaussian form is defined around each grid
point that satisfies the initial conditions given in equation
(\ref{eq:tegainicond}). For the calculations presented in this paper
200 evenly spaced grid points in the range $(-8,8)$ is 
sufficient for converging the configuration space matrix elements of the
Boltzmann operator in the range $(-6,6)$. For calculating time dependent
averages as in equation (\ref{eq:defave}), Gauss-Hermite quadrature is used.

To obtain the configuration matrix elements of the Boltzmann
operator $\exp(-\beta {\hat H})$, the equations of motion, equations
(\ref{eq:tegaeom1})-(\ref{eq:tegaeom3}), were integrated up to the half time
$\hbar\beta/2$ using the adaptive step size Cash-Karp Runge-Kutta
method \cite{NRC}.
  The matrices of the TEGA
propagator and the correction operator are calculated and stored at
every time step. Then, the higher order terms in the series
corresponding to half time are calculated recursively by using equation
(\ref{eq:kiexp}). Finally, by using the symmetric formula,
equation (\ref{eq:symmattr}), the matrix elements of the higher
order terms in the series expansion of the Boltzmann operator are
calculated to find the matrix elements corresponding to the full time.

The calculations are done at three different temperatures. Their results 
are given below. 
\subsubsection{Symmetric Potential, $T=2000\, \mathrm{^o K}$}

\begin{figure}[!tbp]
\begin{center}
\input{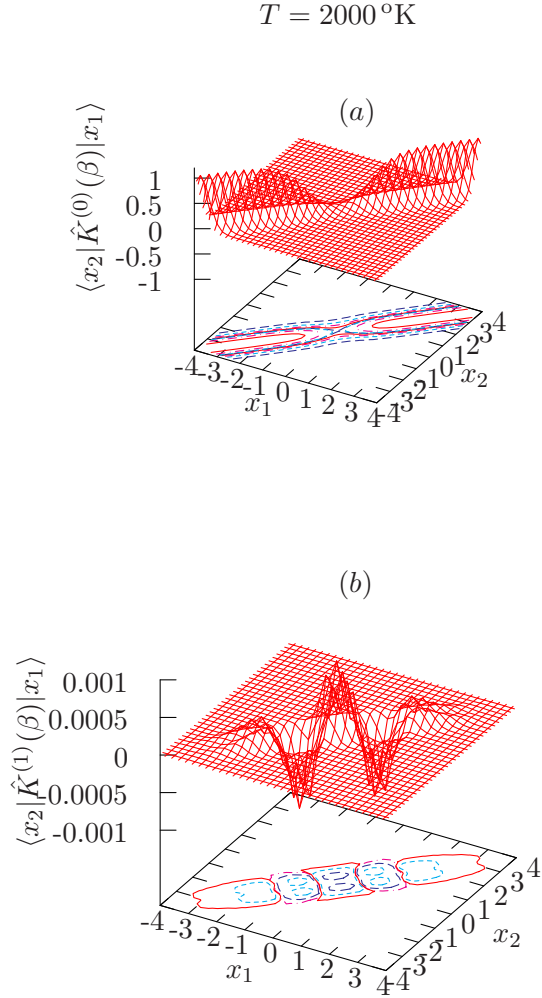}
\end{center}
\caption{\label{fig:k2000} 
3-D plots and contour plots of the first two terms in the
series expansion of the thermal propagator at temperature $T=2000\,
\mathrm{^o K}$ for the symmetric Eckart barrier potential.}
\end{figure}

\begin{figure}[!tbp]
\begin{center}
\input{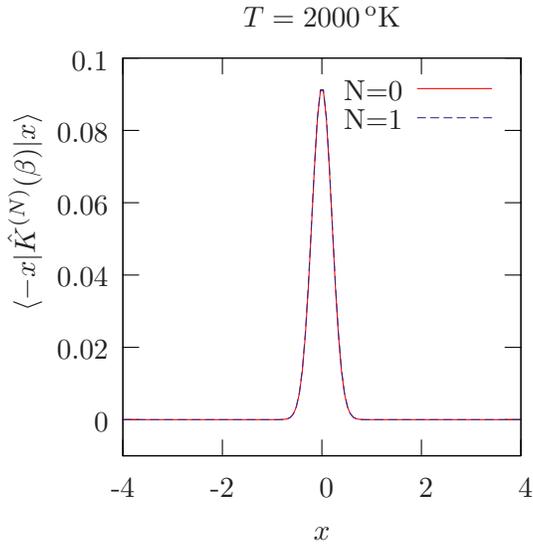}
\end{center}
\caption{One dimensional cuts along the anti-diagonal of the matrix
 elements of the thermal propagator at temperature
$T=2000\, \mathrm{^o K}$ for the symmetric Eckart barrier potential.
  \label{fig:dia2000}}
\end{figure}

In the high temperature limit, the Boltzmann operator is well
approximated in terms of classical mechanics. Since the TEGA reduces
to the classical mechanical Boltzmann distribution, one expects it
to be accurate in this limit.
Figure \ref{fig:k2000} shows surface and contour plots of the
matrix elements of the terms $\hat{K}^{(0)}(\beta)$ and
$\hat{K}^{(1)}(\beta)$ at the temperature $T=2000\, \mathrm{^oK}$. In reduced
variables, $\hbar\beta\omega^{\ddag}=1.20$ which is small when
compared to the reduced crossover temperature of $2\pi$ between
thermal activation and tunneling ($\omega^{\ddag}$ is the harmonic
barrier frequency of the Eckart barrier). From the figure one notes
that the first order term in the series expansion is indeed very
small as compared to the zeroth order term. 

Since it is difficult to quantitatively compare contour plots, 
in figure \ref{fig:dia2000}, a cut of the contour plot along the
antisymmetric line $x'=-x$ is shown. One notes from the figure, that there is
virtually no difference between the zero-th and first order results.

\subsubsection{Symmetric Potential, $T=200\, \mathrm{^o K}$}
\begin{figure*}[!tbp]
\begin{center}
\input{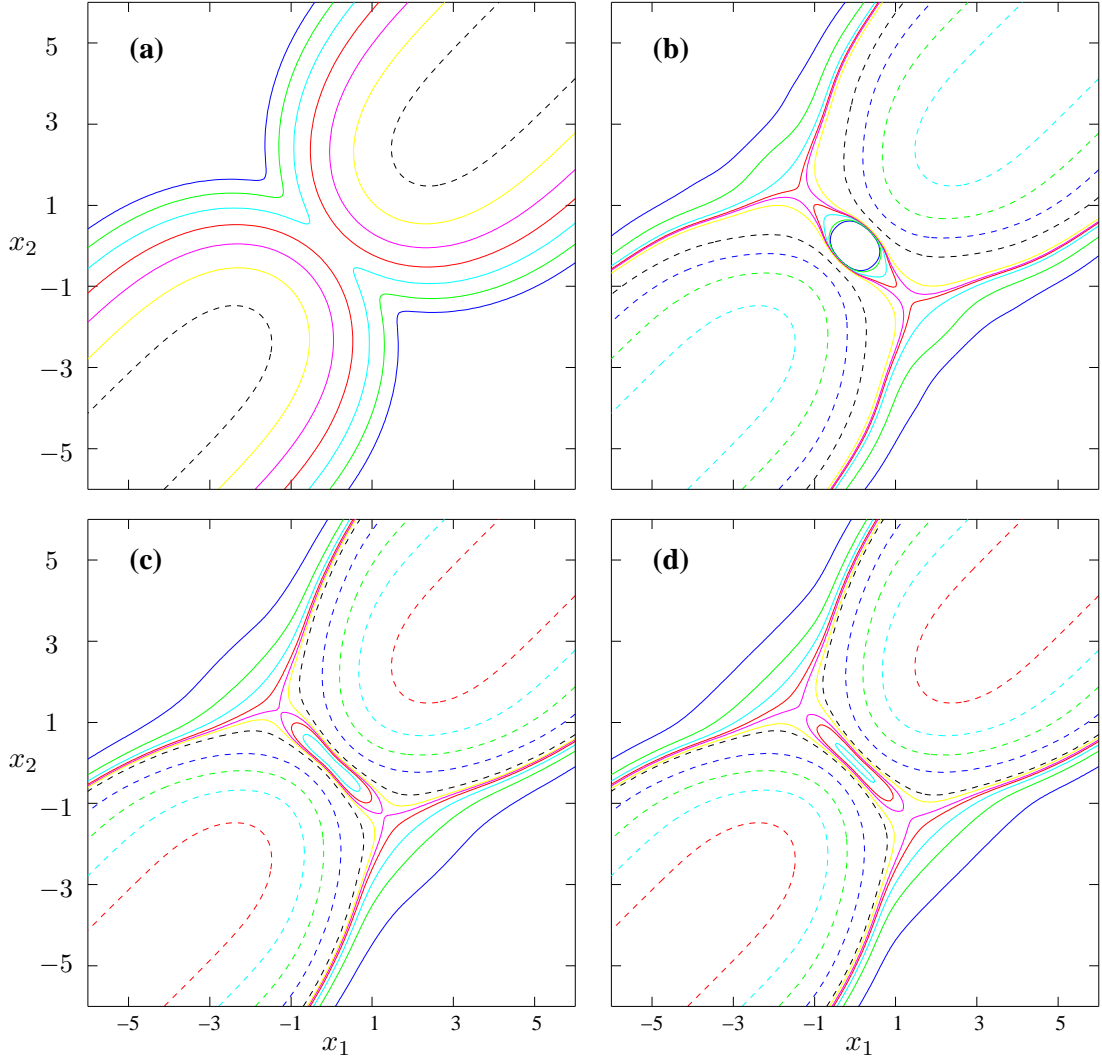}
\end{center}
\caption{Contour plots of the matrix elements of the thermal
propagator in the TEGA series expansion. Panels $(a)$, $(b)$, $(c)$ and
$(d)$ correspond to the truncated series expansion of the
 order $N=0,1,2,3$ respectively at
temperature $T=200\, \mathrm{^o K}$ for the symmetric Eckart barrier
potential. In panel $(a)$, the contour values are $10^{-n}$ with $n=1,\ldots,7$.
The highest contour ($n=1$) is the solid black line, and the lowest contour
($n=7$) is the dark blue line. In panel $(b)$ the contour values are 
$10^{-1}$(dashed light blue line), $10^{-2},10^{-3},10^{-4},10^{-5}$(solid 
yellow line), $7.0\times 10^{-6},6.0\times10^{-6}$(solid red line), 
$4.0\times10^{-6}$(solid light blue line),$10^{-6}$, $10^{-7}$(solid dark blue line). 
The contour values for panels $(c)$ and $(d)$ are $10^{-1}$(dashed red line),
$10^{-2}$, $10^{-3}$, $10^{-4}$, $10^{-5}$, 
$6.0\times10^{-6}$(solid yellow line), $5.0\times10^{-6}$,
 $4.0\times 10^{-6}$(solid red line), $2.5 \times 10^{-6}$(solid light
blue line), $10^{-6}, 
10^{-7}$(solid dark blue line).
\label{fig:k200}}
\end{figure*}

\begin{figure}[!tbp]
\begin{center}
\input{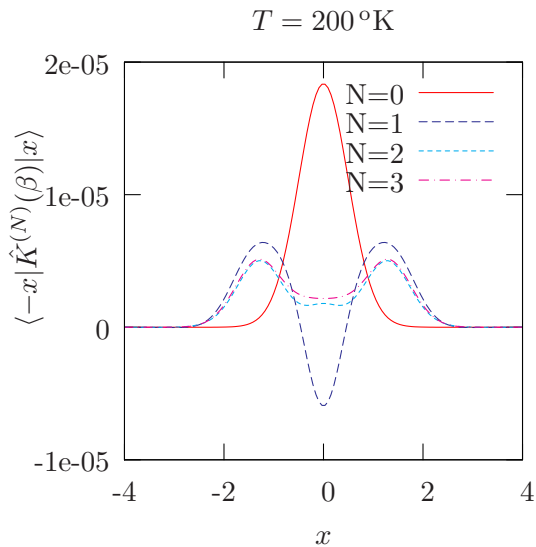}
\end{center}
\caption{One dimensional cuts along the anti-diagonal of 
 the matrix elements of the thermal propagator are shown for
the truncated series up to order $3$ at temperature $T=200\,
\mathrm{^o K}$ for the symmetric Eckart barrier potential.
\label{fig:dia200}}
\end{figure}

Figure \ref{fig:k200} shows  contour plots of the matrix elements
of the imaginary time propagator $\sum_{j=0}^{j=N}
\hat{K}^{(j)}(\beta)$, for $N=0,1,2,3$, respectively at the temperature
$T=200\, \mathrm{^oK}$ 
(or $\hbar\beta\omega^{\ddagger}=12$) which is below the (reduced)
crossover temperature of $2\pi$. At this temperature, tunneling
becomes important so that the zero-th order contribution in the
series representation is no longer sufficient. The contour plot of
the matrix elements of the TEGA (zero-th order) propagator as shown
in panel (a) of the Figure has a single saddle point structure as
predicted by Liu and Miller \cite{Liu2006}. 
As shown in panel (b)
of figure \ref{fig:k200}, adding the first order term changes the structure
of the contour plot completely such that there are two saddle points in the
graph instead of one.  From the figure, it is seen that to obtain the
numerically exact results, it is necessary to include terms up to
order $3$. Convergence of the results can also be followed from the matrix 
elements $\langle -x \vert \hat{K}^{N}(\beta)\vert x \rangle$
that are plotted in figure \ref{fig:dia200}.

\subsubsection{Symmetric Potential, $T=40\, \mathrm{^o K}$}
\begin{figure}[!tbp]
\begin{center}
\input{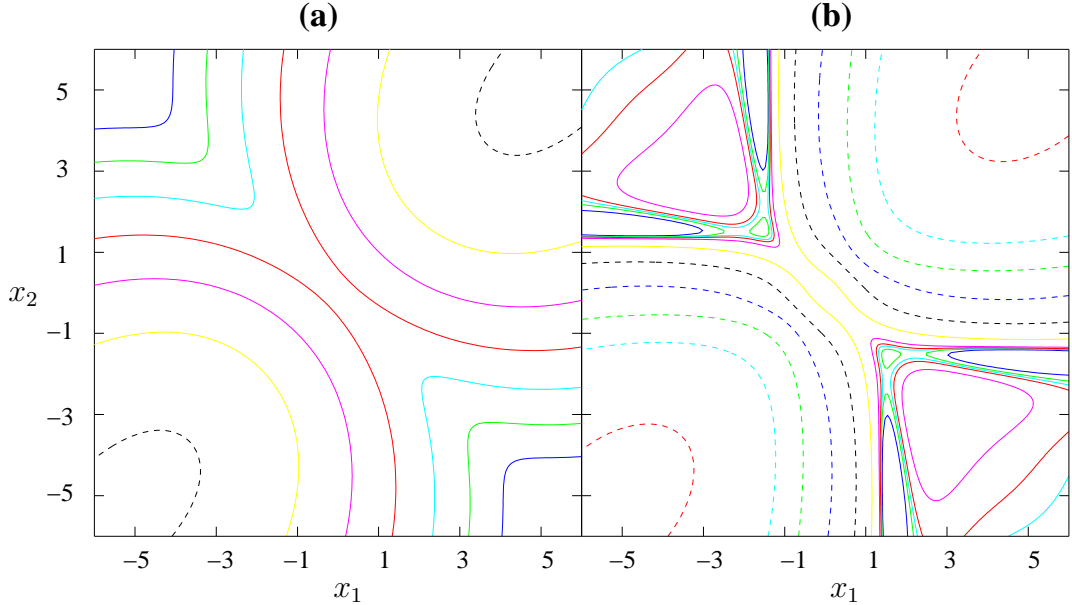}
\end{center}
\caption{Contour plots of the matrix elements of the zeroth and
fifth order truncated series representation of  the thermal
propagator at the temperature $T=40\, \mathrm{^o K}$ for the
symmetric Eckart barrier potential. In panel $(a)$, the contour values are
$10^{-n}$ with $n=1,\ldots,7$. The highest contour (n=1) is the black dashed
 line, and the lowest contour (n=7) is the dark blue line. 
In panel $(b)$, the contour values are 
$10^{-1}$(dashed red line), $10^{-2}$, $10^{-3}$, $10^{-4}$, $10^{-5}$, 
$10^{-6}$(solid yellow line), $3.0\times 10^{-7}$, $10^{-7}$,
$10^{-8}$, $-10^{-8}$(solid green line),
$-5.0\times 10^{-8}$(solid dark blue line).
\label{fig:k40}}
\end{figure}

\begin{figure}[!tbp]
\begin{center}
\begin{tabular}{c}
$T=40 \, \mathrm{^o K}$ \\
\input{dia-40loc.tex} \\
\input{dia-40hoc.tex}
\end{tabular}
\end{center}
\caption{One dimensional cuts along the anti-diagonal of 
the matrix elements of the thermal propagator at
temperature $T=40\, \mathrm{^o K}$ for the symmetric Eckart barrier
potential.
\label{fig:dia40}}
\end{figure}

The reduced temperature when $T=40\,\mathrm{^oK}$ is 
$\hbar\beta\omega^{\ddagger}=60$, which is 
much larger than $2\pi$, so that this temperature corresponds to 
 ``deep'' tunneling
regime. As expected, when the temperature is lowered, the
calculations become more demanding. The converged results are
obtained only after including the fifth order term in the series
expansion of the imaginary time propagator. The convergence can be 
followed from figure \ref{fig:dia40} where one dimensional cuts along the
antisymmetric line for the different terms in the series are plotted.
Contour plots are given in figure \ref{fig:k40}.

Interestingly, the structure becomes even more
complicated. Along the antisymmetric line, there are three maxima.
The saddle point reappears at the origin and there are four additional
saddle points which are off of the antisymmetry line. This has
implications for the quantum instanton method \cite{miller2003},
where the two dividing surfaces are taken along the two saddle
points, as found at the higher temperature. Presumably, one could
still take the two dividing surfaces to be at the point on the
antisymmetric line obtained from the intersection of the
antisymmetric line and the line that connects each pair of saddle
points. However, this needs to be studied in more detail.

\subsubsection{Asymmetric Potential, $T=200\, \mathrm{^o K}$}
\begin{figure}[!tbp]
\begin{center}
\input{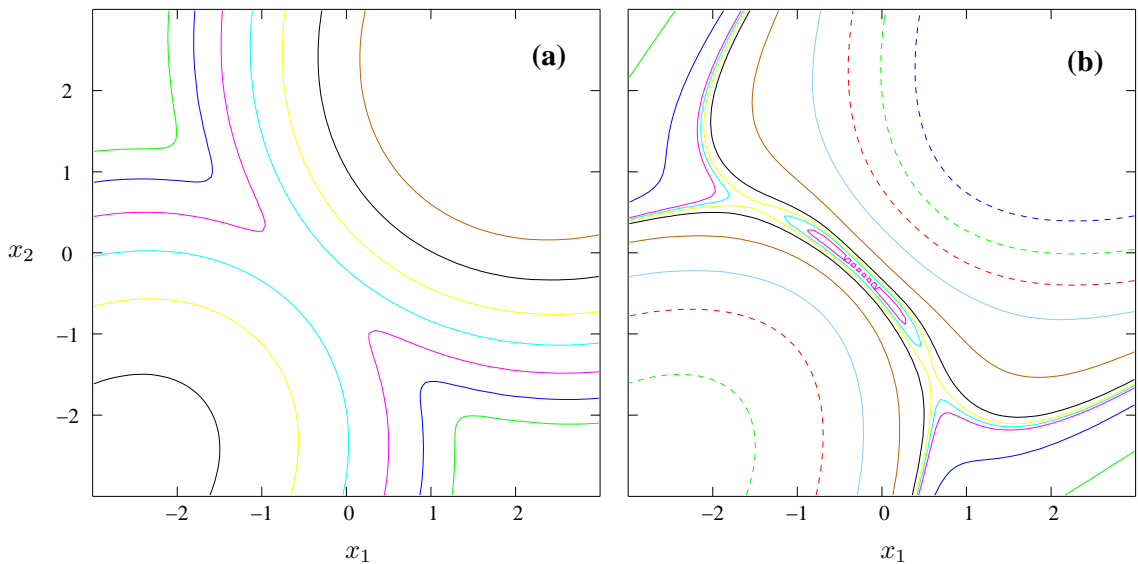}
\end{center}
\caption{Contour plots of the matrix elements of the  zeroth and fifth
order truncated series representation of the thermal propagator
at temperature $T=200\, \mathrm{^o K}$ for the asymmetric
Eckart barrier potential. In panel $(a)$, the contour values are $10^{-n}$
with $n=0,\ldots6$. The highest contour ($n=0$) 
appears only on the upper right corner of the graph, and the lowest contour
($n=6$) is the green line. 
In panel $(b)$, the contour values are $1$, $10^{-1}$, $10^{-2}$,
$10^{-3}$, $10^{-4}$, 
$3.0\times10^{-5}$ ,
 $2.5\times 10^{-5}$(yellow line), $2.3\times 10^{-5}$(solid light blue line), 
$2.1\times 10^{-5}$, $10^{-5}$(solid dark blue line), $10^{-6}$.
 \label{fig:a200}}
\end{figure}

The results of the TEGA calculation for the asymmetric potential are
shown in figure \ref{fig:a200}. Here too, it was necessary to
include all terms up to fifth order for convergence. 
Due to the
asymmetry of the barrier, a plot of the matrix elements 
$\langle -x \vert \hat{K}^{(N)}(\beta)\vert x\rangle$ do not show any 
structure. Besides, it is not helpful for following the convergence of the
results, either. For these reasons, they are not plotted. 
In this case, the reduced temperature 
$\hbar\beta\omega^{\ddag}=12.5$ is below the crossover temperature,
however not very much lower so that again one has only two saddle
points. Contour plots of matrix elements for the zeroth order approximation
and the converged results are given in figure \ref{fig:a200}.

\subsubsection{\label{ssec:comp}A Comparison Of The Results
 With The Results Of Miller {\it et. al.}}

\begin{figure}
\begin{center}
\input{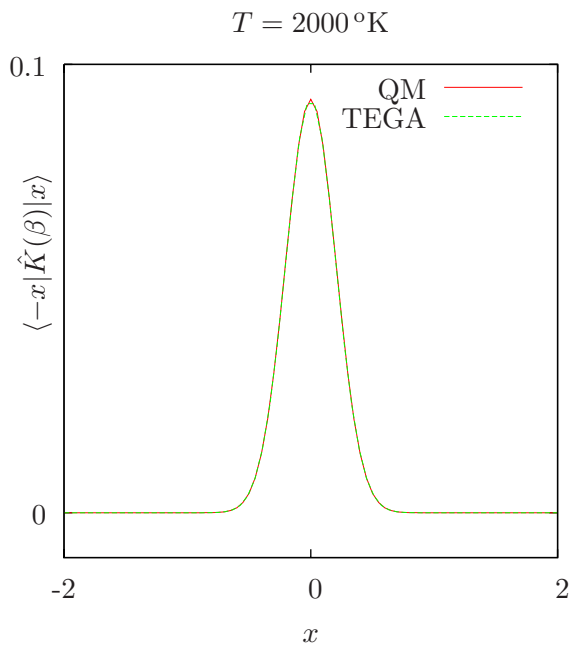}
\caption{\label{fig:compdia2000}  A comparison of the results of quantum
mechanical and the converged TEGA calculations of the matrix elements 
$\langle -x \vert \hat{K}(\beta) \vert x \rangle$ for the symmetric Eckart
barrier potential at temperature $T= 2000 \, \mathrm{^o K}$ .}
\end{center}
\end{figure}

\begin{figure}
\begin{center}
\input{comp-dia-200.tex}
\caption{\label{fig:compdia200}A comparison of the results of quantum 
mechanical and the converged TEGA calculations of the matrix elements 
$\langle -x \vert \hat{K}(\beta) \vert x \rangle$ for the symmetric Eckart
barrier potential at temperature $T= 200 \, \mathrm{^o K}$.}
\end{center}
\end{figure}

\begin{figure}
\begin{center}
\input{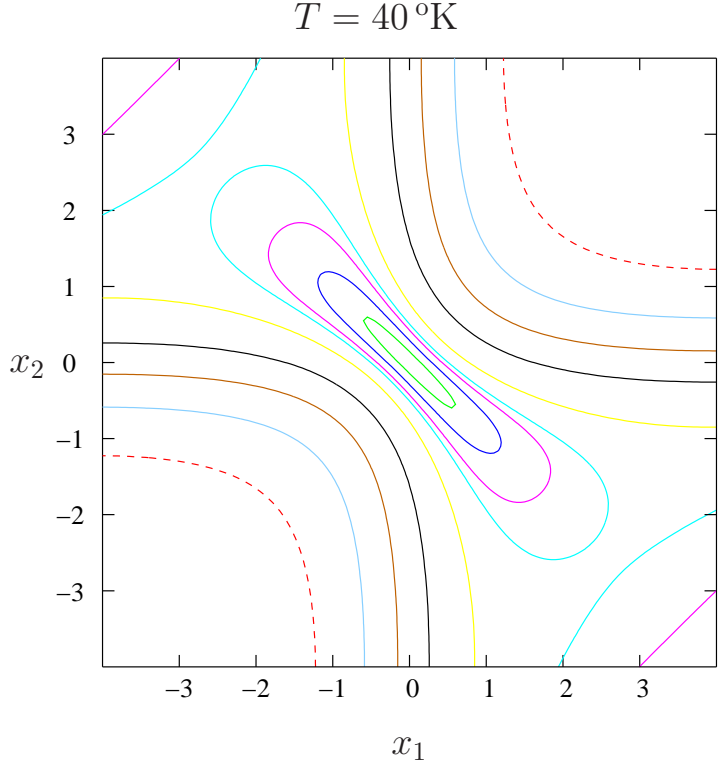}
\caption{\label{fig:qdmat40}Contour plot of the results of the quantum
mechanical calculation of the matrix elements 
$\langle x_2 \vert \exp(-\beta \hat{H})\vert x_1\rangle$ for the 
symmetric Eckart barrier potential at temperature 
$T=40\, \mathrm{^o K}$.
 In the figure, contour values are $10^{-2}$(dashed red line),
$10^{-3}$, $10^{-4}$, $10^{-5}$, $10^{-6}$, $3\times 10^{-7}$(solid light blue
line), $2\times 10^{-7}$, $10^{-7}$, $5 \times 10^{-8}$(solid green line).}
\end{center}
\end{figure}

\begin{figure}
\begin{center}
\input{comp-dia-40.tex}
\caption{\label{fig:compdia40}A comparison of the results of quantum mechanical
and the converged TEGA calculations of the matrix elements 
$\langle -x \vert \hat{K}(\beta)\vert x \rangle $ for the symmetric Eckart 
barrier potential at temperature $T=40 \, \mathrm{^o K}$.}
\end{center}
\end{figure}

\begin{figure}
\begin{center}
\input{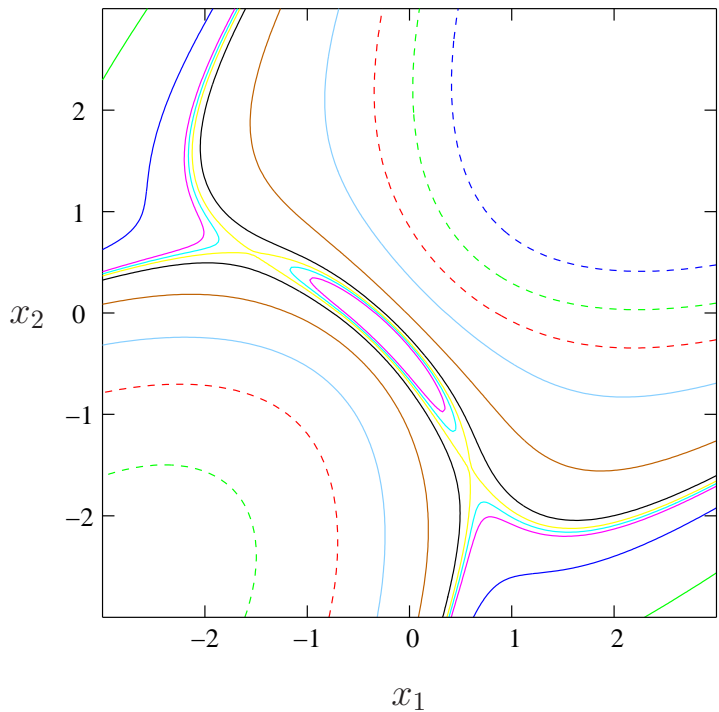}
\caption{\label{fig:qdmata200} Contour plot of the results of quantum mechanical
calculation of the matrix elements 
$\langle x_2 \vert \exp (-\beta \hat{H})\vert x_1 \rangle $ for the asymmetric
Eckart barrier potential at temperature $T=200\, \mathrm{^o K}$. The contour values
are the same with the ones that are used in figure \ref{fig:a200}.}
\end{center}
\end{figure}

The matrix elements of the Boltzmann operator for the symmetric and 
asymmetric Eckart barrier had been calculated previously by Miller 
{\it et. al.}. However, in that paper, authors did not provided 
contour values in the plots. For this reason, their calculations are
duplicated. First, the Hamiltonian matrix 
is diagonalized in a sinc DVR basis \cite{Colbert91}. Then, the configuration
matrix elements of the Boltzmann operator are calculated by 
using equation \ref{eq:qdenmat}.
 In the diagonalization calculation, grid spacing is
taken to be $0.05$a.u. where the grid ranges from $-20$a.u. to $20$a.u.  

For the results of the calculations at temperatures $T=2000\, \mathrm{^oK}$ and
$T=200\,\mathrm{^oK}$, when the contour plots of the results of the 
quantum mechanical calculations were prepared (with the same contour values
that are used in the contour plots of TEGA calculations),
 there was no visual difference
between them and the contour plots of the TEGA calculations so that they are
not given. The results of the quantum mechanical and the TEGA calculations
of the matrix elements $\langle -x \vert \hat{K}(\beta)\vert x \rangle$
are compared in figures \ref{fig:compdia2000} and \ref{fig:compdia200}
for the temperatures $T=2000\,\mathrm{^o K}$ and 
$T=200\,\mathrm{^o K}$, respectively. 
From figure \ref{fig:compdia2000}, it can be seen that the agreement between
the quantum mechanical results and the TEGA results is perfect at 
temperature $T=2000\,\mathrm{ ^o K}$.
 At temperature $T=200\, \mathrm{^o K}$, agreement is still
quite good, but the results differ a little bit in the tunneling region. 

On the other hand, when the results of two different calculations are
compared for $T=40\, \mathrm{^o K}$, they differ both quantitively and 
qualitatively. The contour plot of the results of quantum mechanical
calculation is given in figure \ref{fig:qdmat40}. In the contour plot,
there are still two saddle points as in the $T=200\, \mathrm{^o K}$ case.
However, the TEGA calculation predict more than two saddle points at 
that temperature. The difference of the results can also be seen from
figure \ref{fig:compdia40} where the results of the two calculations are
given for the matrix elements $\langle -x \vert \hat{K}(\beta)\vert x \rangle$.
The difference in the results can be attributed to the fact that the way that
the quantum
mechanical calculation is performed is not a proper way of calculating the
matrix elements of the imaginary time propagator especially at such low 
temperatures. Because, in this calculations one is imposing artificial 
infinite potential walls at the boundaries which does not make sense 
if the potential surface of the system does not support bound states. 
Since the Eckart potential has a continuous spectrum and do not have any bound 
states, introduction of these artificial infinite walls is a source of
error, because it discritizes a continuous system. The errors 
might be expected to be small at high temperatures, which can also
be seen from the comparisons of the results at temperatures 
$T=2000\, \mathrm{^o K}$ and $T=200\, \mathrm{^o K}$.
 However, the errors can be significant at low temperatures.
Miller {\it et. al.}
did not make these calculations at $T=40\, \mathrm {^o K}$ anyway.
 Furthermore, as will
be seen in section \ref{ssec:dblwl} 
increasing number of saddle points is again observed in double well calculations
both with proper quantum mechanical calculations and also with TEGA and 
PSTEGA 
calculations.

Finally, a quantum mechanical calculation of the matrix elements of the 
Boltzmann operator for the asymmetric Eckart barrier potential at temperature
$T=200\, \mathrm{^o K}$ results in 
very good agreement with the TEGA calculations. 
The contour plot of the results of quantum mechanical calculation is 
given in figure \ref{fig:qdmata200}.

\subsection{\label{ssec:dblwl}Double Well Potential}

Calculations are done with a one dimensional
double well potential surface 
which has the form 
\begin{equation}
V(x) = V_0 (ax^4+bx^2+c),
\end{equation}
where $V_0=0.004 \mathrm{a.u.}$, $a= 1.0 (\mathrm{a.u.})^{-4}$, 
$b=-4.0 (\mathrm{a.u.})^{-2}$, $c=4.0$. 
The mass of the particle is taken
to be $m=1061.0 \mathrm{a.u.}$. The calculations are done by using
both the bare potential and also the Gaussian averaging. Three different
temperatures are used in the calculations which are: $T=2000\, \mathrm{^oK}$, 
$T=400\, \mathrm{^oK}$, and $T=100\, \mathrm{^oK}$. 
In order to calculate the matrix elements of the density matrix, 
the same numerical procedure that was used in the
Eckart Barrier calculations is followed.
Firstly, an equally spaced grid of $300$ points in the range [-3:3]
is taken. Then, the Coherent States, that are formed around these grid points
are propagated up to half time using the adaptive step size 
Cash-Karp Runge-Kutta method \cite{NRC}. The matrices of the zeroth order
approximations to the propagator and the correction operator are stored 
in every time step. Then, the matrices of the higher order terms in the 
series expansion of the propagator is obtained by using the recursion 
formula (equation \ref{eq:kiexp}). Time integrations are performed with
third order Simpson's integration. Finally, by using the symmetric formula,
equation (\ref{eq:symexp}), matrix elements corresponding to full time 
is calculated. Please note that, Gaussian averages of the potential and 
its derivatives can be calculated analytically for the potential surface 
studied here. 
Quantum mechanical calculations are done with the same parameters
given in section \ref{ssec:comp} for the Eckart barrier 
calculations. The results that are obtained with the TEGA and
the PSTEGA methods will be presented in the following subsections. 
Before that, it should be noted that all of the 
results in this section are scaled with the partition function which is 
$Z= \mathrm{Tr}[\exp(-\beta \hat{H})]$.

\subsubsection{$T=2000\, \mathrm{^o K}$}

\begin{figure}
\begin{center}
\input{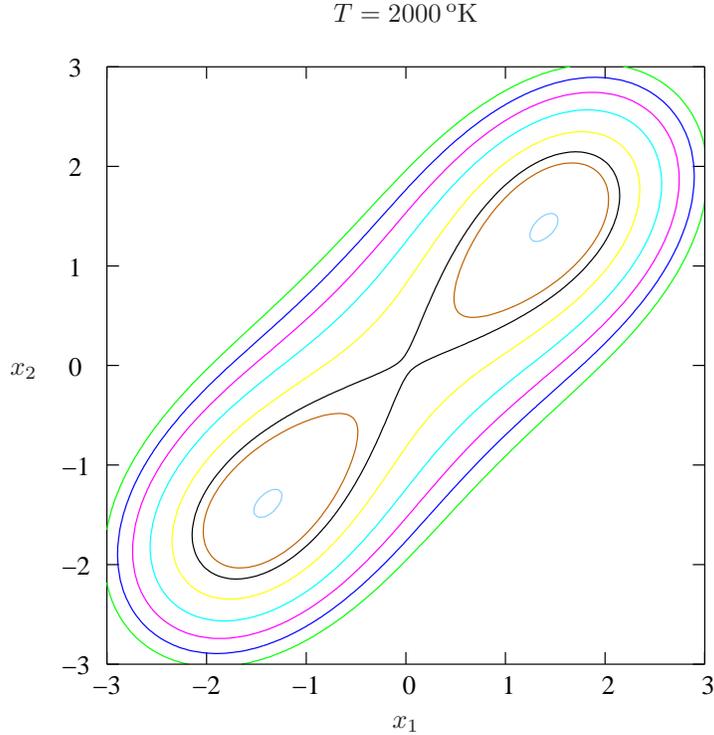}
\end{center}
\caption{\label{fig:dmat2000}
A contour plot of the matrix elements of the Boltzmann operator
at temperature
 $T=2000 \mathrm{^o K}$. All of the different calculations gives almost
identical results so that the same graph is obtained from all of them.
 The contour values are: 
$5 \times 10^{-1}, 10^{-1}, 5 \times 10^{-2}, 10^{-2}, 10^{-3}, 10^{-4},10^{-5},$ and $10^{-6}$.}
\end{figure}

\begin{figure}
\begin{center}
\input{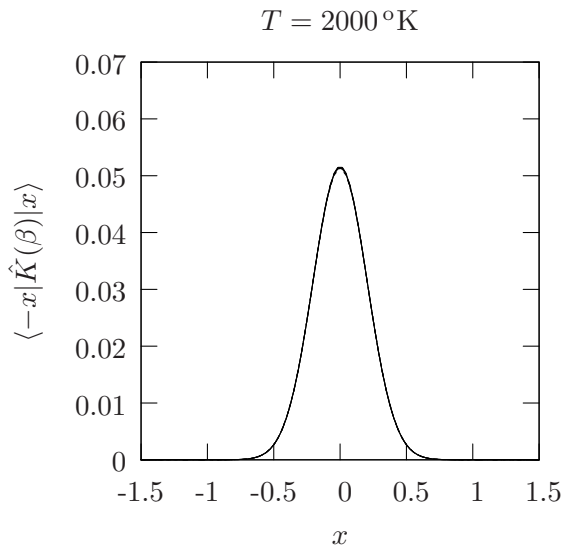}
\end{center}
\caption{\label{fig:ad2000}
One dimensional cuts along the anti-diagonal of the matrix elements
of the Boltzmann operator is shown at temperature $T=2000 \mathrm{^o K}$. 
In the figure there exists five different plots (results of both TEGA and PSTEGA
calculations calculated with both the Gaussian averaged potential and the 
bare potential, and also the results of quantum mechanical calculation).
 However, there is no way to differentiate between them.}
\end{figure}

In figure \ref{fig:dmat2000}, a contour plot of the matrix elements of the 
Boltzmann operator at temperature $T=2000 \, \mathrm{^o K}$ is shown.
 That figure is prepared by using the results
of the zeroth order 
TEGA calculation with the averaged potential. The results of the
other calculations, including the quantum mechanical calculation, 
give almost the identical results so that it was not necessary to prepare
different graphs for different calculations. This can be realized easily
from figure \ref{fig:ad2000}, in which the matrix elements 
$\langle -x \vert \exp (- \beta \hat{H}) \vert x \rangle$ are plotted for all 
of the five different calculations. As it can be seen from the figure,
there is no way to differentiate between different graphs. 
Therefore, at high temperatures use of the bare potential in propagation
does not  lead to a worse zeroth order approximation approximation 
than the use of the Gaussian averaged potential. This can be 
attributed to the fact that such high temperatures involves very short
time propagation so that the Gaussians which are very narrow initially do 
not get much broadened. Therefore, use of such narrow Gaussians for 
averaging is like using delta functions, which is the same
thing with using bare potentials. Consequently, use of the bare potential
for such short propagation times does not lead to a significant error in the
zeroth order approximations.  

\subsubsection{$T=400\, \mathrm{^o K}$}

\begin{figure}
\begin{center}
\input{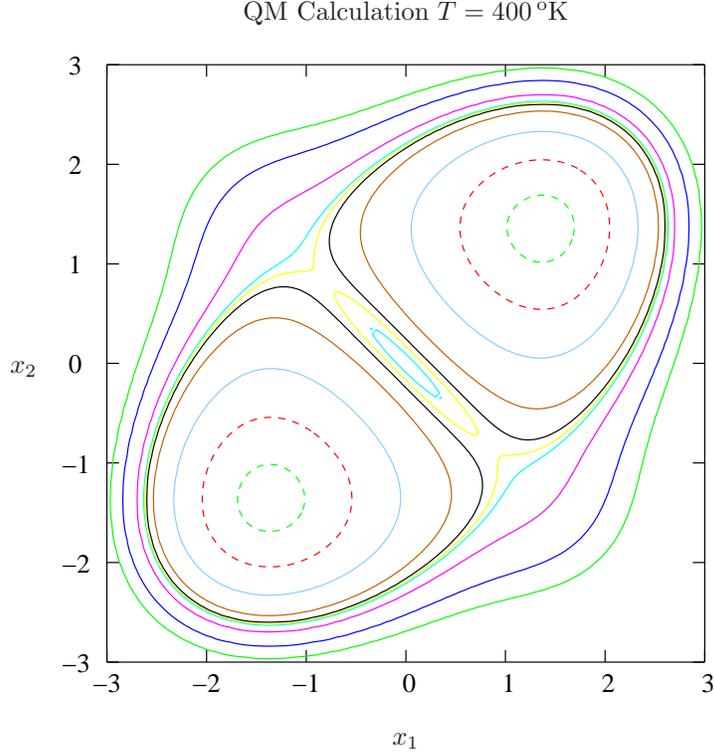}
\end{center}
\caption{\label{fig:qdmat400}
A contour plot of the matrix elements of the thermal propagator
obtained with a quantum mechanical calculation at temperature 
$T=400\,  \mathrm{^o K}$. The contour values are the same with the ones
that are used in figure \ref{fig:dmat400}.}
\end{figure}

\begin{figure*}
\begin{center}
\input{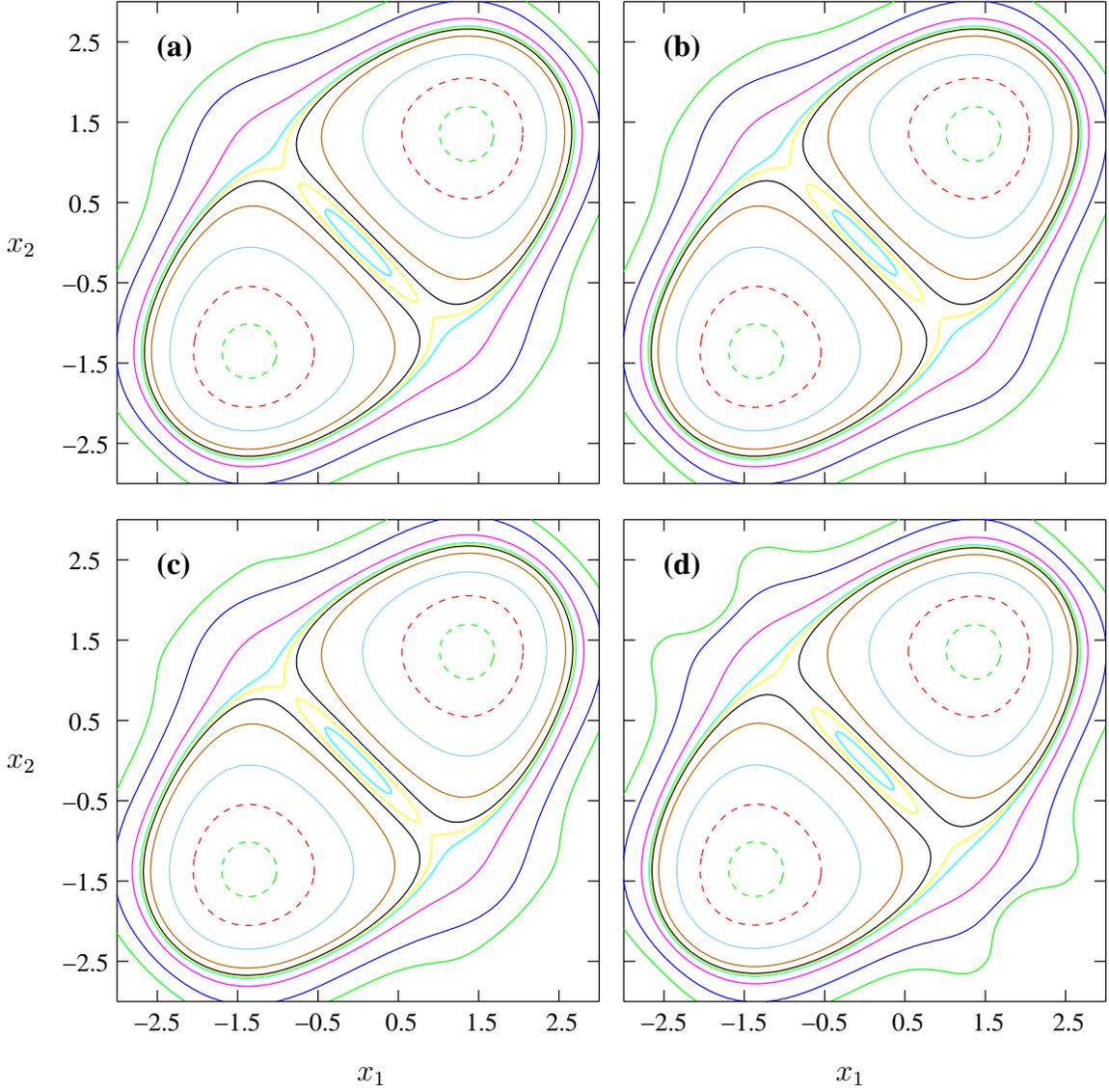}
\end{center}
\caption{\label{fig:dmat400}
Contour plots of the matrix elements of the thermal propagator 
at temperature $T=400\, \mathrm{^o K}$ is
shown in panels $(a)$, $(b)$, $(c)$ and $(d)$ for TEGA calculation with the 
Gaussian averaged potential , 
PSTEGA calculation with the Gaussian averaged potential, TEGA
calculation with the bare potential and PSTEGA calculation with the bare
potential, respectively. The results of the Gaussian averaged calculations
refer to truncated series of order $3$ while the results of the 
bare potential calculations refer to truncated series of order $4$.
Time step that is used in bare potential calculations was half of the
time step that is used in Gaussian averaged calculations. The contour
values are: $5\times 10^{-1}$(dashed green line),
 $10^{-1}$, $10^{-2}$, $10^{-3}$, $4\times 10^{-4}$,
$3.2\times10^{-4}$(solid yellow line), $2.7 \times 10^{-4}$(solid light blue
line), $10^{-4}$, $10^{-5}$ and $10^{-6}$(solid green line).}
\end{figure*}

\begin{figure*}
\begin{center}
\input{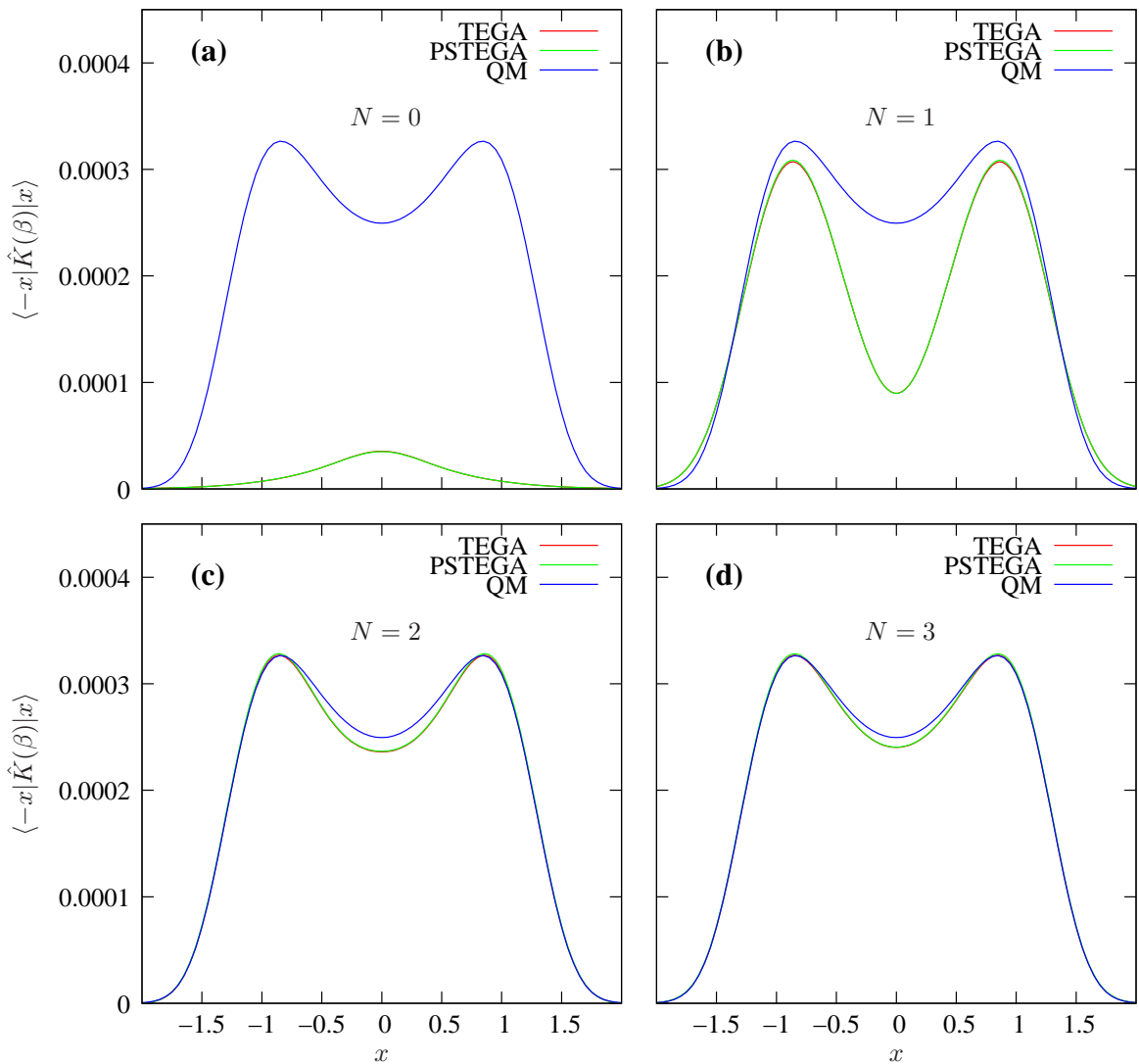}
\end{center}
\caption{\label{fig:ad400}
One dimensional cuts along the anti-diagonal of the matrix elements
of the thermal propagator is shown at temperature $T=400\,\mathrm{^o K}$ 
for TEGA and PSTEGA calculations with the Gaussian averaged potential.
In the figure, panels $(a)$, $(b)$, $(c)$ and $(d)$ refer to the results
of the truncated series of order $N=0,1,2$ and $3$, respectively. Results of
quantum mechanical calculations are also shown in each panel.}
\end{figure*}

\begin{figure*}
\begin{center}
\input{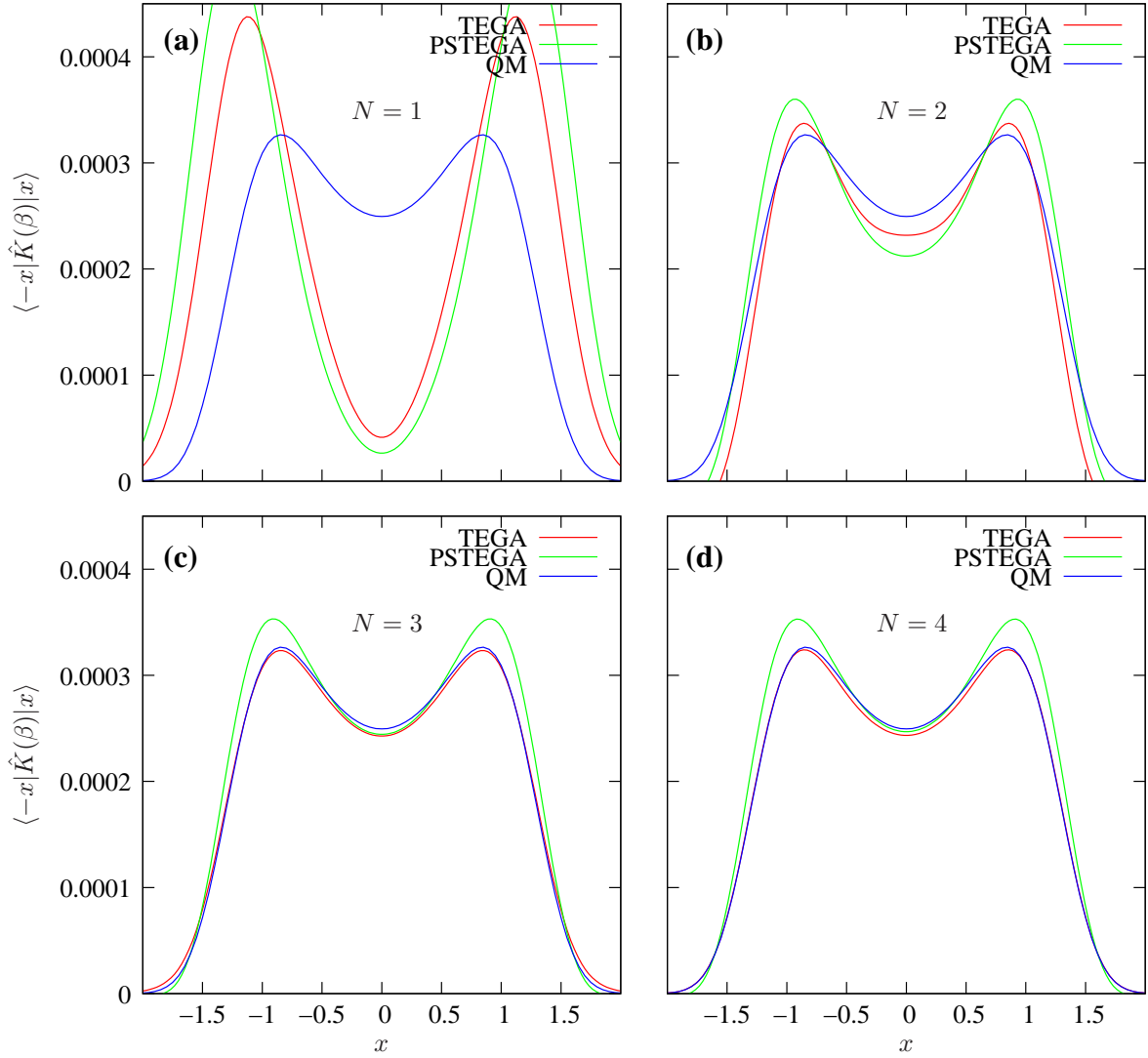}
\end{center}
\caption{\label{fig:adb400}
One dimensional cuts along the anti-diagonal of the matrix elements of 
the thermal propagator at temperature $T=400\, \mathrm{^o K}$. The results of
TEGA and  PSTEGA calculations are obtained by using the bare potential.
The panels $(a)$, $(b)$, $(c)$ and $(d)$ refer to the results of the 
truncated series of order $N=0,1,2$, and $3$, respectively. The results of quantum 
mechanical calculations are also shown in each panel. }
\end{figure*}

At temperature $T=400\, \mathrm{^o K}$, tunneling becomes important so that
the zeroth order approximations to the thermal propagator does not 
lead to accurate results. In fact, as it is shown before \cite{Liu2006}
they do not even give the correct qualitative picture since they always lead
to a single saddle point at (0,0).  Nevertheless, use of the series 
representation of the propagator converges to the right answers. In order
to make a comparison of the TEGA and the PSTEGA calculations, the results
of the quantum mechanical calculation is shown in figure \ref{fig:qdmat400}.
Converged results of the TEGA and PSTEGA calculations are shown in 
figure \ref{fig:dmat400}. For the Gaussian averaged potential 
calculations , it was 
necessary to include terms up to third order. Convergence of the 
Gaussian averaged potential calculations are shown for the matrix elements
$\langle -x \vert  K^{(i)}(\beta)\vert x\rangle$ in figure \ref{fig:ad400}.
It can be seen from the figure that the zeroth order approximation 
leads to quite bad results both qualitatively and quantitatively. 
Inclusion of the first order correction gives accurate results in 
some parts of the configuration space, but not in the tunneling region.
In order to get accurate results also in the tunneling region, it is
necessary to include even the third order correction terms. 

For the bare potential calculations, it was necessary to include even the
forth order correction terms in order to converge the results. Besides,
it was necessary to use smaller time steps in time integrations. Contour
plots of the converged results of the TEGA and PSTEGA calculations are
shown in figure \ref{fig:dmat400}. In figure \ref{fig:adb400}, convergence
of the results is shown for the matrix elements
$\langle -x \vert  K^{(i)}(\beta)\vert x\rangle$ for $i=0,\ldots,3$. Comparing the results
in this figure with the results obtained with Gaussian averaged potential
calculations, shown in figure \ref{fig:ad400}, it can be seen that the 
use of the bare potential  leads to much lower accuracy in the zeroth order
approximations. Nevertheless, the calculations converge to quite accurate
results. Besides, despite the fact that fluctuations in the results 
for the bare potential calculations are much bigger than the fluctuations
in the results for the Gaussian averaged calculations, it can be 
said that the series representations show similar convergence properties
for both of the calculations since one of them converges at the third order
and the other at the fourth order.

\subsubsection{$T=100\, \mathrm{^o K}$}
\begin{figure*}
\begin{center}
\input{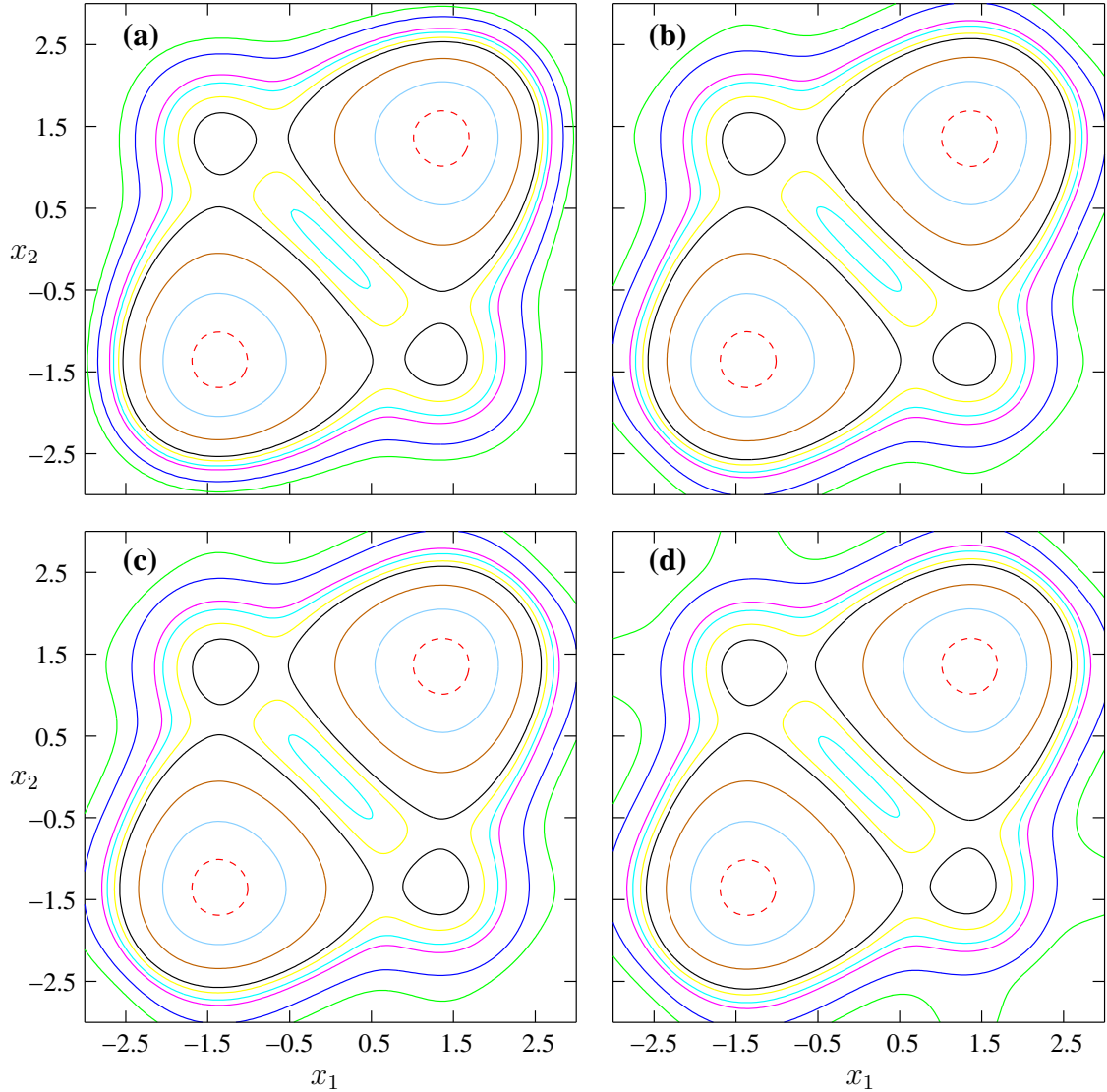}
\end{center}
\caption{\label{fig:dmat100}
Contour plots of the matrix elements of the Boltzmann operator 
at temperature $T=100\, \mathrm{^o K}$. The results of quantum mechanical
calculation, TEGA calculation with the Gaussian averaged potential, 
PSTEGA calculation with the Gaussian averaged potential, and TEGA
calculation with the bare potential are shown in panels $(a)$, $(b)$
$(c)$ and $(d)$, respectively. Results of the Gaussian averaged calculations
refer to truncated series of order $3$ while the result of the bare
potential calculation refers to truncated series of order $4$.
The contour values are: $5\times 10^{-1}$(dashed red line),
 $10^{-1}, 10^{-2}$, $10^{-3}$,
$5 \times 10^{-4}$(solid yellow line),
 $2 \times 10^{-4}$(solid light blue line),
 $10^{-4}$, $10^{-5}$ and $10^{-6}$.}
\end{figure*}

\begin{figure*}
\begin{center}
\input{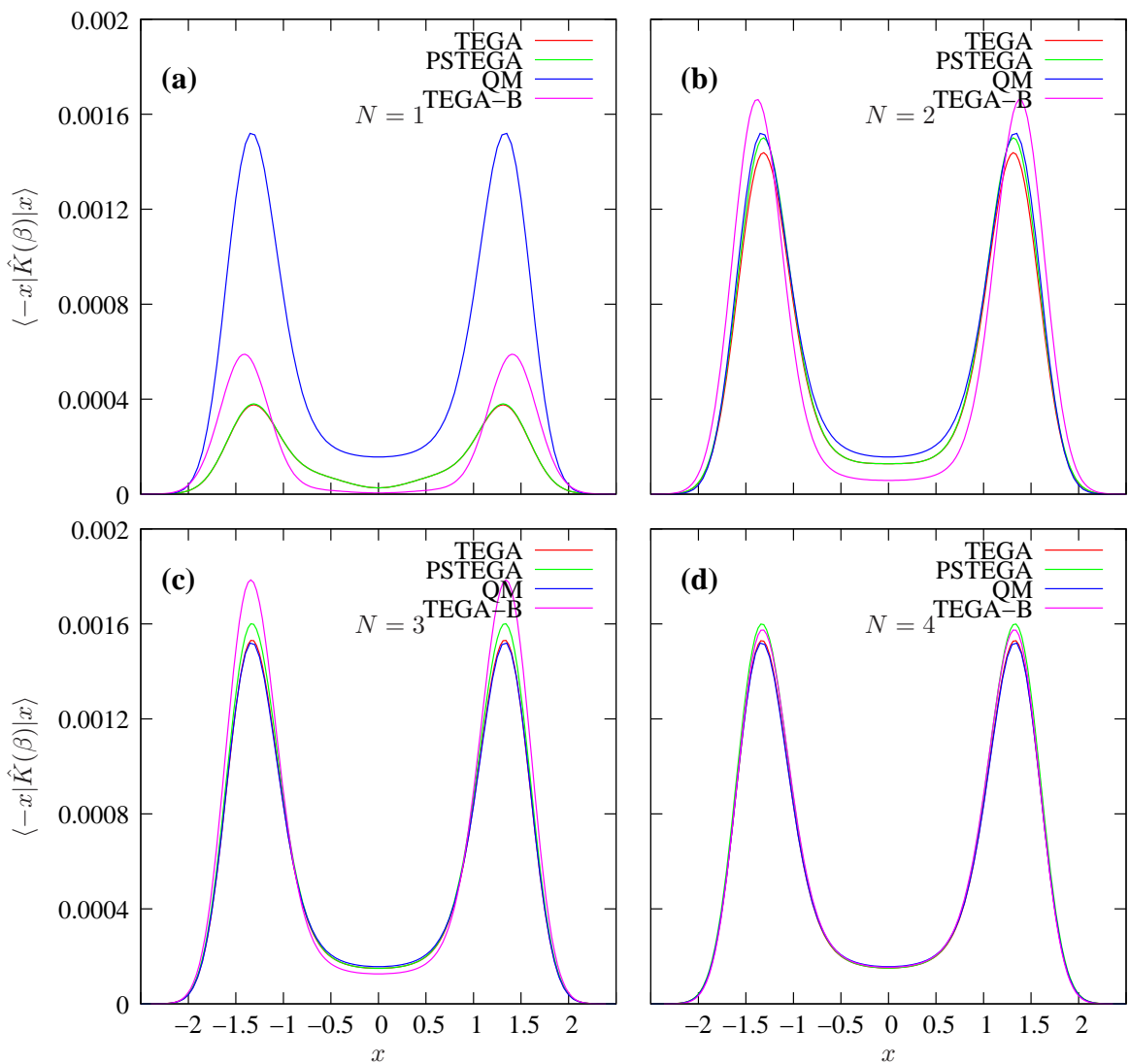}
\end{center}
\caption{\label{fig:ad100}
One dimensional cuts along the anti-diagonal of the matrix elements
of the thermal propagator at temperature $T=100\, \mathrm{^o K}$. The panels
$(a)$, $(b)$, $(c)$ and $(d)$ refer to truncated series of order $N=1,2,3$ and
$4$, respectively. 
Results of  quantum mechanical calculations are also shown
in each panel. }
\end{figure*}

At temperature $T=100 ^o \mathrm{K}$, tunneling becomes even more important.
Besides, the calculations get more demanding because of 
the increasing propagation
time. 

A contour plot of the
 results of quantum mechanical calculation is shown in panel $(a)$ of figure
\ref{fig:dmat100}. Contour plots of the results of TEGA and PSTEGA calculations
with the Gaussian averaged potentials is shown in panels $(b)$ and $(c)$ of the
same figure. At this temperature, the series for 
the PSTEGA calculation with the bare potential
surface does not converge. On the other hand, the series for the 
TEGA calculation  with the bare potential surface still converges. A contour
plot of the results of the TEGA calculation with the bare potential is shown
in panel $(d)$ of 
figure \ref{fig:dmat100}. All of the graphs looks very similar.
In order to converge the series, it was necessary to include terms up to
order 3 for the Gaussian averaged potential surface calculations and terms up
to order 4 for the bare potential calculations. 

Convergence of the results can again be followed from the antisymmetric line.
In figure \ref{fig:ad100}, one dimensional cuts along the anti-diagonal 
of the matrix elements of the thermal propagator is shown for the truncated 
series of order $N=1,2,3,4$ in panels $(a)$, $(b)$, $(c)$ and $(d)$ 
respectively. The results of quantum mechanical calculation is also shown
in each panel. 

Another thing which needs to be noted about the contour plots is the presence
of more than two saddle points. As in the Eckart Barrier calculations,
 it is again observed
that the saddle points move away from the antisymmetric line. From figure
\ref{fig:dmat100}, it can be seen that 
there exists four saddle points.

\subsubsection{A Discussion Of The Results}
In Eckart Barrier calculations, it was observed that the results of the 
quantum mechanical calculations do not agree with the
results of TEGA calculations at low temperatures.
It was argued that the discrepancy between the TEGA and the quantum mechanical
results should be related with the artificial discretization of 
a continuous system by imposition of wrong boundary conditions
to quantum mechanical calculations. On the other hand,
double well potential surface has a discrete spectrum and 
do not support any scattering states. Therefore, the bound state 
calculation is a proper way of performing 
a quantum mechanical calculation for calculating the matrix elements of 
the equilibrium density matrix. The agreement of the results of 
quantum mechanical 
calculation with the results of the TEGA and PSTEGA calculations 
are very good in this case at all temperatures.  This also supports that
the reason of the discrepancy in the Eckart barrier 
calculations is related with 
the imposition of wrong boundary conditions to quantum mechanical calculations.

Considering the zeroth order TEGA and PSTEGA approximations, their accuracy
depends on the temperature. At high temperatures, where the system is 
almost classical, zeroth order approximations lead to accurate
results. As the 
temperature is lowered, accuracy of the zeroth order approximations gets
worse as expected since the quantum effects becomes important at low
temperatures. As shown by Liu and Miller \cite{Liu2006}, TEGA always leads
to a single saddle point at (0,0). A similar analysis can also be made for
PSTEGA and it can be shown that it is also the case for PSTEGA. Therefore,
both TEGA and PSTEGA do not even lead to correct structure at 
low temperatures where the tunneling effects are important.
Nevertheless, if the series expansion converges, both of them converge to 
the correct answers even at low temperatures. 

Use of the bare potential results in lower accuracy in the zeroth order 
approximation compare to use of the Gaussian averaged potential. Higher 
accuracy of the results of Gaussian averaged calculations can be attributed
to the fact that Gaussian averaging of the potential surface results from
variational principles. Nevertheless, use of the bare potential leads to 
faster integration of equations of motion. However, it also leads to 
slower convergence of the series expansion. Besides, the results fluctuate
more during convergence if the bare potential is used. 
In this study, it was not possible to converge the results at 
$100\, \mathrm{^o K}$ with the PSTEGA method if the calculations are done with
the bare potential. On the other hand, if Gaussian averaged potential
is used, the series expansion for the PSTEGA method still converges, and 
it gives accurate results at that temperature. 

One thing needs to be noted about PSTEGA calculations. While integrating
the equations of motion initial width of the Gaussian wave packet is 
arbitrary. However, this does not mean that one can take any value for the
initial width and converge the calculations to the correct results. While 
doing PSTEGA computations, it was necessary to figure out which initial
width gives the best answers. This is done by comparing the results of 
PSTEGA calculations with the results of TEGA calculations. It was seen that 
if the initial width of the Gaussian is taken to be $\approx 1$ 
(in mass weighted coordinates), then the 
results of TEGA and the PSTEGA methods are almost identical for the 
Gaussian averaged calculations. 
This is true for both the zeroth order approximations and also for the
truncated series of any order. In other words, $n^{\mathrm{th}}$ order
PSTEGA expansion and $n^{\mathrm{th}}$ order TEGA expansion gives identical
results within numerical accuracy if the initial arbitrary width of the 
Gaussians in PSTEGA calculations is chosen good. 

\section{\label{sec:conc}Discussions and Conclusions}

The TEGA and the PSTEGA
series representations of the thermal propagator were tested
for two different potential surfaces. The results show that
the number of terms needed in the series increases as the
temperature is lowered. However, even for a reduced temperature as
low as $\hbar\beta\omega^{\ddag}=60$ the expansion converges by the
time one reaches the fifth order in the series. In real time, this
would make the computation prohibitive, since it would be impossible
to converge such high order terms using Monte Carlo methods for a
multidimensional system. In imaginary time, the integrand is much
less oscillatory and so there is hope that even when dealing with
many degrees of freedom, one could converge the higher order terms.

Even if it turns out that it is not practical to converge the higher
order terms of the series when the system is ``complex'', there is
value in the present computation. It does show that the series
converges rather rapidly and that the series at least in principle
does lead to the correct result.


In this paper, numerical convergence properties of the TEGA and
the  PSTEGA series
representations of the imaginary time propagator are compared. It is shown
that if the initial arbitrary width of the Gaussians are chosen good;
then, TEGA and PSTEGA methods gives identical results within numerical 
accuracy. Although, the PSTEGA method involves a phase space integration,
it is shown in the appendix that the momentum coordinates can be integrated
implicitly, so that the PSTEGA method can also be implemented in configuration
space. Thus, in both the TEGA and the PSTEGA methods, number of equations
of motion scales linearly with the dimension of the problem. 

It is seen that the Gaussian averaging is important especially at low 
temperatures. The use of the bare potential leads to very low accuracy 
for the zeroth order term of the series representation such that it causes
the series representation not to converge at low temperatures.   

Another important thing which puts a challenge to semiclassical analysis 
is that it is observed that the number of the saddle points of the matrix
elements $\langle x'\vert \hat{K}(\beta) \vert x\rangle $ increases as 
the temperature is lowered. 
Semiclassically, it is obvious why one
should expect two saddle points. As analyzed by Miller {\it et. al.}
\cite{miller2003}, the two saddle points correspond semiclassically
to the two turning points of the classical periodic orbit on the
upside down potential energy surface whose half period is
$\hbar\beta$ \cite{MillerTST}. However, it is 
found that as the temperature is lowered,
additional saddle points show up. These point out the need for
perhaps a deeper semiclassical analysis at low temperature. They
also create a challenge to the quantum instanton method which used
the two saddle points to identify the relevant dividing surfaces for
thermal rate computations. At the low temperatures, at which one
finds more than two saddle points, it is not clear which saddle
points should be used within the quantum instanton method context.
This question may become even more acute when dealing with
asymmetric systems.

\newpage
\appendix

\section*{An Efficient Way of Integrating 
	Equations of Motion for PSTEGA Calculations \label{sec:app}}

Equation (\ref{eq:peom}) can be integrated implicitly to give 
\begin{equation}
	\BS{p}(\tau)= \BS{c}(\tau) \BS{p}(0),
\end{equation}
where $\BS{c}(\tau)$ is given by 
\begin{equation}
	\BS{c}(\BS{q},\tau)= \exp \left( - \int_0^{\tau} \mathrm{d}\tau' 
		\hbar^2 \BS{G}(\tau')^{-1}\right),
\end{equation}
with the initial condition $\BS{c}(\BS{q},0)=I$,
 which can be integrated with the 
equation of motion 
\begin{equation}
\frac{\partial \BS{c}(\BS{q},\tau)}{\partial \tau} = - \hbar^2 \BS{G}(\tau)^{-1}
	\BS{c}(\BS{q},\tau). 
\end{equation}
It is useful to define some 
auxiliary equations of motion that helps to integrate Gaussian integrals of
$\BS{p}(\tau)$. 
With the following definitions:
\begin{eqnarray}
\BS{k}(\tau) & = &
 \int_0^{\tau} \mathrm{d}\tau'\BS{c}(\BS{q},\tau')^T \BS{c}(\BS{q},\tau') \\
\BS{w}(\BS{q},\tau) & = & \exp \left( - \int_0^{\tau} \mathrm{d} \tau' 
	\langle V(\BS{q}(\tau')) \rangle + \frac{\hbar^2}{4} 
		\mathrm{Tr}[\BS{G}(\tau')^{-1}] \right) \label{eq:wdef} \\
\BS{s}(\BS{q},\tau) & = & \frac{1}{\hbar}
	\int_0^{\tau} \mathrm{d}\tau' \BS{c}(\BS{q},\tau')
	\BS{G}(\tau') \langle \nabla V (\BS{q}(\tau')) \rangle 
\end{eqnarray}
and integrating the following equations of motion,
\begin{eqnarray}
\frac{\partial \BS{k}(\BS{q},\tau)}{\partial \tau} & =  &
	\BS{c}(\BS{q},\tau)^T \BS{c}(\BS{q},\tau),
	\; \; \BS{k}(\BS{q}_0,0)=\BS{0}, \\
\frac{\partial \BS{w}(\BS{q},\tau)}{\partial \tau} & = & 
	-\BS{w}(\BS{q},\tau) \left( \langle V(\BS{q}(\tau))\rangle 
		 + \frac{\hbar^2}{4} 
		\mathrm{Tr}[\BS{G}(\tau)^{-1}] 
	\right), \; \; \BS{w}(\BS{q}_0,0)=I, \label{eq:weom}\\
\frac{\partial \BS{s}(\BS{q},\tau)}{\partial \tau} & = &  
	\BS{c}(\BS{q},\tau)^T \BS{G}(\tau) 
	\langle \nabla V(\BS{q}(\tau))\rangle, \; \;
	\BS{s}(\BS{q}_0,0) = \BS{0},  
\end{eqnarray}
matrix elements of the zeroth order approximation to the propagator  can be
obtained as 
\begin{eqnarray}
\langle \BS{x} \vert \hat{K}_0(\tau) \vert \BS{x}' \rangle & = &
	\int \frac{\mathrm{d}\BS{p} \mathrm{d}\BS{q}}{2 \pi} 
	\langle \BS{x} \vert \hat{K}_0(\tau) \vert g(\BS{p},\BS{q},0) \rangle 
	\langle g(\BS{p},\BS{q},0) \vert \BS{x}'\rangle  \\
& = &  \int \frac{\mathrm{d}\BS{q}}{2 \pi} \frac{(2 \pi)^{N/2}}{\sqrt{\mathrm{det}(\BS{k}(\BS{q},\tau))}} \BS{w}(\BS{q},\tau) \left( \frac{1}{\mathrm{det}(\BS{G}(\tau)\BS{G}(0))} \right)^{1/4} l(\BS{q},\BS{x},\BS{x}',\tau),
\end{eqnarray}
where 
\begin{eqnarray}
l(\BS{q},\BS{x},\BS{x}',\tau)& = & \exp (- \frac{1}{2}(
	 \BS{t}(\BS{q},\tau)^T \BS{k}(\BS{q},\tau)^{-1}\BS{t}(\BS{q},\tau)
	+ (\BS{x}-\BS{q}(\tau))^T \BS{G}(\tau)^{-1}(\BS{x}-\BS{q}(\tau))
	 \nonumber \\ 
 & & 	+ (\BS{x}'-\BS{q}_0)\BS{G}(0)^{-1}(\BS{x}'-\BS{q}_0)) ),
\end{eqnarray}
where 
\begin{equation}
\BS{t}(\BS{q},\tau) = \frac{1}{\hbar}\BS{s}(\BS{q},\tau) 
	+ \frac{1}{\hbar} \BS{c}(\BS{q},\tau)(\BS{x}-\BS{q}(\tau))
	- \frac{1}{\hbar}(\BS{x}'-\BS{q}_0).
\end{equation}
If the calculations are done with the 
bare potential, the following term should be
added to the expression in parenthesis in equations (\ref{eq:wdef}) and
(\ref{eq:weom}),
\begin{equation}
	 \frac{1}{4} \mathrm{Tr}[\nabla \nabla^T V(\BS{q}(\tau))\BS{G}(\tau)];
\end{equation}
and also the Gaussian averagings of the potential 
and its derivatives are not performed. 

Although, the integration scheme described above increases the number of
equations of motion per particle.
 It reduces the phase space integration to a configuration
space integration so that the total number of equations of motion is 
greatly reduced. 

\newpage
\bibliography{bibliography}

\begin{thebibliography}{34}
\expandafter\ifx\csname natexlab\endcsname\relax\def\natexlab#1{#1}\fi
\expandafter\ifx\csname bibnamefont\endcsname\relax
  \def\bibnamefont#1{#1}\fi
\expandafter\ifx\csname bibfnamefont\endcsname\relax
  \def\bibfnamefont#1{#1}\fi
\expandafter\ifx\csname citenamefont\endcsname\relax
  \def\citenamefont#1{#1}\fi
\expandafter\ifx\csname url\endcsname\relax
  \def\url#1{\texttt{#1}}\fi
\expandafter\ifx\csname urlprefix\endcsname\relax\def\urlprefix{URL }\fi
\providecommand{\bibinfo}[2]{#2}
\providecommand{\eprint}[2][]{\url{#2}}

\bibitem[{\citenamefont{{Van Vleck}}(1928)}]{Vleck28}
\bibinfo{author}{\bibfnamefont{J.~H.} \bibnamefont{{Van Vleck}}},
  \bibinfo{journal}{Proc. Nat. Ac. Sci. USA} \textbf{\bibinfo{volume}{14}},
  \bibinfo{pages}{178} (\bibinfo{year}{1928}).

\bibitem[{\citenamefont{Gutzwiller}(1971)}]{Gutzwiller}
\bibinfo{author}{\bibfnamefont{M.~C.} \bibnamefont{Gutzwiller}},
  \bibinfo{journal}{J. Math. Phys.} \textbf{\bibinfo{volume}{12}},
  \bibinfo{pages}{343} (\bibinfo{year}{1971}).

\bibitem[{\citenamefont{Miller}(1970)}]{Miller70}
\bibinfo{author}{\bibfnamefont{W.~H.} \bibnamefont{Miller}},
  \bibinfo{journal}{J. Chem. Phys.} \textbf{\bibinfo{volume}{53}},
  \bibinfo{pages}{3578} (\bibinfo{year}{1970}).

\bibitem[{\citenamefont{Marcus}(1971)}]{Marcus71}
\bibinfo{author}{\bibfnamefont{R.~A.} \bibnamefont{Marcus}},
  \bibinfo{journal}{J. Chem. Phys.} \textbf{\bibinfo{volume}{54}},
  \bibinfo{pages}{3965} (\bibinfo{year}{1971}).

\bibitem[{\citenamefont{Heller}(1975)}]{Heller75}
\bibinfo{author}{\bibfnamefont{E.~J.} \bibnamefont{Heller}},
  \bibinfo{journal}{J. Chem. Phys.} \textbf{\bibinfo{volume}{62}},
  \bibinfo{pages}{1544} (\bibinfo{year}{1975}).

\bibitem[{\citenamefont{Heller}(1981)}]{Heller81}
\bibinfo{author}{\bibfnamefont{E.~J.} \bibnamefont{Heller}},
  \bibinfo{journal}{J. Chem. Phys.} \textbf{\bibinfo{volume}{75}},
  \bibinfo{pages}{2923} (\bibinfo{year}{1981}).

\bibitem[{\citenamefont{Herman and Kluk}(1984)}]{Herman84}
\bibinfo{author}{\bibfnamefont{M.~F.} \bibnamefont{Herman}} \bibnamefont{and}
  \bibinfo{author}{\bibfnamefont{E.}~\bibnamefont{Kluk}},
  \bibinfo{journal}{Chem. Phys.} \textbf{\bibinfo{volume}{91}},
  \bibinfo{pages}{27} (\bibinfo{year}{1984}).

\bibitem[{\citenamefont{Herman}(1994)}]{Herman94}
\bibinfo{author}{\bibfnamefont{M.~F.} \bibnamefont{Herman}},
  \bibinfo{journal}{Annu. Rev. Phys. Chem.} \textbf{\bibinfo{volume}{45}},
  \bibinfo{pages}{83} (\bibinfo{year}{1994}).

\bibitem[{\citenamefont{{Sep{\'u}lveda} and Grossman}(1996)}]{Grossman96}
\bibinfo{author}{\bibfnamefont{M.~A.} \bibnamefont{{Sep{\'u}lveda}}}
  \bibnamefont{and} \bibinfo{author}{\bibfnamefont{F.}~\bibnamefont{Grossman}},
  \bibinfo{journal}{Adv. Chem. Phys.} \textbf{\bibinfo{volume}{96}},
  \bibinfo{pages}{191} (\bibinfo{year}{1996}).

\bibitem[{\citenamefont{Miller}(1997)}]{Miller97}
\bibinfo{author}{\bibfnamefont{W.~H.} \bibnamefont{Miller}},
  \bibinfo{journal}{Adv. Chem. Phys.} \textbf{\bibinfo{volume}{101}},
  \bibinfo{pages}{853} (\bibinfo{year}{1997}).

\bibitem[{\citenamefont{Miller}(1998)}]{Miller98}
\bibinfo{author}{\bibfnamefont{W.~H.} \bibnamefont{Miller}},
  \bibinfo{journal}{Faraday Disc.} \textbf{\bibinfo{volume}{110}},
  \bibinfo{pages}{1} (\bibinfo{year}{1998}).

\bibitem[{\citenamefont{Tannor and Garashchuk}(2000)}]{Tannor2000}
\bibinfo{author}{\bibfnamefont{D.~J.} \bibnamefont{Tannor}} \bibnamefont{and}
  \bibinfo{author}{\bibfnamefont{S.}~\bibnamefont{Garashchuk}},
  \bibinfo{journal}{Ann. Rev. Phys. Chem.} \textbf{\bibinfo{volume}{51}},
  \bibinfo{pages}{553} (\bibinfo{year}{2000}).

\bibitem[{\citenamefont{Baranger et~al.}(2001)\citenamefont{Baranger, {de
  Aguiar}, Keck, Korsch, and Schellhaass}}]{Baranger2001}
\bibinfo{author}{\bibfnamefont{M.}~\bibnamefont{Baranger}},
  \bibinfo{author}{\bibfnamefont{M.~A.~M.} \bibnamefont{{de Aguiar}}},
  \bibinfo{author}{\bibfnamefont{F.}~\bibnamefont{Keck}},
  \bibinfo{author}{\bibfnamefont{H.~J.} \bibnamefont{Korsch}},
  \bibnamefont{and}
  \bibinfo{author}{\bibfnamefont{B.}~\bibnamefont{Schellhaass}},
  \bibinfo{journal}{J. Phys. A: Mat. Gen.} \textbf{\bibinfo{volume}{34}},
  \bibinfo{pages}{7227} (\bibinfo{year}{2001}).

\bibitem[{\citenamefont{Miller}(2001)}]{Miller2001}
\bibinfo{author}{\bibfnamefont{W.~H.} \bibnamefont{Miller}},
  \bibinfo{journal}{J. Phys. Chem. A} \textbf{\bibinfo{volume}{105}},
  \bibinfo{pages}{2942} (\bibinfo{year}{2001}).

\bibitem[{\citenamefont{Thoss and Wang}(2004)}]{Thoss2004}
\bibinfo{author}{\bibfnamefont{M.}~\bibnamefont{Thoss}} \bibnamefont{and}
  \bibinfo{author}{\bibfnamefont{H.}~\bibnamefont{Wang}},
  \bibinfo{journal}{Annu. Rev. Phys. Chem.} \textbf{\bibinfo{volume}{55}},
  \bibinfo{pages}{299} (\bibinfo{year}{2004}).

\bibitem[{\citenamefont{Kay}(2005)}]{Kay2005}
\bibinfo{author}{\bibfnamefont{K.~G.} \bibnamefont{Kay}},
  \bibinfo{journal}{Ann. Rev. Phys. Chem.} \textbf{\bibinfo{volume}{56}},
  \bibinfo{pages}{225} (\bibinfo{year}{2005}).

\bibitem[{\citenamefont{Hellsing et~al.}(1985)\citenamefont{Hellsing, Sawada,
  and Metiu}}]{Metiu85}
\bibinfo{author}{\bibfnamefont{B.}~\bibnamefont{Hellsing}},
  \bibinfo{author}{\bibfnamefont{S.-I.} \bibnamefont{Sawada}},
  \bibnamefont{and} \bibinfo{author}{\bibfnamefont{H.}~\bibnamefont{Metiu}},
  \bibinfo{journal}{Chem. Phys. Lett.} \textbf{\bibinfo{volume}{122}},
  \bibinfo{pages}{303} (\bibinfo{year}{1985}).

\bibitem[{\citenamefont{Makri and Miller}(2002)}]{Makri2002}
\bibinfo{author}{\bibfnamefont{N.}~\bibnamefont{Makri}} \bibnamefont{and}
  \bibinfo{author}{\bibfnamefont{W.~H.} \bibnamefont{Miller}},
  \bibinfo{journal}{J. Chem. Phys.} \textbf{\bibinfo{volume}{116}},
  \bibinfo{pages}{9207} (\bibinfo{year}{2002}).

\bibitem[{\citenamefont{Frantsuzov et~al.}(2003)\citenamefont{Frantsuzov,
  Neumaier, and Mandelshtam}}]{Frantsuzov2003}
\bibinfo{author}{\bibfnamefont{P.}~\bibnamefont{Frantsuzov}},
  \bibinfo{author}{\bibfnamefont{A.}~\bibnamefont{Neumaier}}, \bibnamefont{and}
  \bibinfo{author}{\bibfnamefont{V.~A.} \bibnamefont{Mandelshtam}},
  \bibinfo{journal}{Chem. Phys. Lett.} \textbf{\bibinfo{volume}{381}},
  \bibinfo{pages}{117} (\bibinfo{year}{2003}).

\bibitem[{\citenamefont{Frantsuzov and Mandelshtam}(2004)}]{Frantsuzov2004}
\bibinfo{author}{\bibfnamefont{P.~A.} \bibnamefont{Frantsuzov}}
  \bibnamefont{and} \bibinfo{author}{\bibfnamefont{V.~A.}
  \bibnamefont{Mandelshtam}}, \bibinfo{journal}{J. Chem. Phys.}
  \textbf{\bibinfo{volume}{121}}, \bibinfo{pages}{9247} (\bibinfo{year}{2004}).

\bibitem[{\citenamefont{Pollak and {Martin-Fierro}}(2007)}]{Pollak2007}
\bibinfo{author}{\bibfnamefont{E.}~\bibnamefont{Pollak}} \bibnamefont{and}
  \bibinfo{author}{\bibfnamefont{E.}~\bibnamefont{{Martin-Fierro}}},
  \bibinfo{journal}{J. Chem. Phys.} \textbf{\bibinfo{volume}{126}},
  \bibinfo{pages}{164107} (\bibinfo{year}{2007}).

\bibitem[{\citenamefont{Ankerhold et~al.}(2001)\citenamefont{Ankerhold,
  Saltzer, and Pollak}}]{Ankerhold2002}
\bibinfo{author}{\bibfnamefont{J.}~\bibnamefont{Ankerhold}},
  \bibinfo{author}{\bibfnamefont{M.}~\bibnamefont{Saltzer}}, \bibnamefont{and}
  \bibinfo{author}{\bibfnamefont{E.}~\bibnamefont{Pollak}},
  \bibinfo{journal}{J. Chem. Phys.} \textbf{\bibinfo{volume}{116}},
  \bibinfo{pages}{5925} (\bibinfo{year}{2001}).

\bibitem[{\citenamefont{Pollak and Shao}(2003)}]{Pollak2003}
\bibinfo{author}{\bibfnamefont{E.}~\bibnamefont{Pollak}} \bibnamefont{and}
  \bibinfo{author}{\bibfnamefont{J.}~\bibnamefont{Shao}}, \bibinfo{journal}{J.
  Phys. Chem. A} \textbf{\bibinfo{volume}{107}}, \bibinfo{pages}{7112}
  (\bibinfo{year}{2003}).

\bibitem[{\citenamefont{Zhang and Pollak}(2003{\natexlab{a}})}]{Zhang2003-2}
\bibinfo{author}{\bibfnamefont{S.}~\bibnamefont{Zhang}} \bibnamefont{and}
  \bibinfo{author}{\bibfnamefont{E.}~\bibnamefont{Pollak}},
  \bibinfo{journal}{Phys. Rev. Lett.} \textbf{\bibinfo{volume}{91}},
  \bibinfo{pages}{190201} (\bibinfo{year}{2003}{\natexlab{a}}).

\bibitem[{\citenamefont{Zhang and Pollak}(2003{\natexlab{b}})}]{Zhang2003}
\bibinfo{author}{\bibfnamefont{S.}~\bibnamefont{Zhang}} \bibnamefont{and}
  \bibinfo{author}{\bibfnamefont{E.}~\bibnamefont{Pollak}},
  \bibinfo{journal}{J. Chem. Phys.} \textbf{\bibinfo{volume}{119}},
  \bibinfo{pages}{11058} (\bibinfo{year}{2003}{\natexlab{b}}).

\bibitem[{\citenamefont{Zhang and Pollak}(2004)}]{Zhang2004}
\bibinfo{author}{\bibfnamefont{S.}~\bibnamefont{Zhang}} \bibnamefont{and}
  \bibinfo{author}{\bibfnamefont{E.}~\bibnamefont{Pollak}},
  \bibinfo{journal}{J. Chem. Phys.} \textbf{\bibinfo{volume}{121}},
  \bibinfo{pages}{3384} (\bibinfo{year}{2004}).

\bibitem[{\citenamefont{Saltzer and Pollak}(2005)}]{Saltzer2005}
\bibinfo{author}{\bibfnamefont{M.}~\bibnamefont{Saltzer}} \bibnamefont{and}
  \bibinfo{author}{\bibfnamefont{E.}~\bibnamefont{Pollak}},
  \bibinfo{journal}{J. Chem. Theo. Comp.} \textbf{\bibinfo{volume}{1}},
  \bibinfo{pages}{439} (\bibinfo{year}{2005}).

\bibitem[{\citenamefont{Shao and Pollak}(2006)}]{Shao2006}
\bibinfo{author}{\bibfnamefont{J.}~\bibnamefont{Shao}} \bibnamefont{and}
  \bibinfo{author}{\bibfnamefont{E.}~\bibnamefont{Pollak}},
  \bibinfo{journal}{J. Chem. Phys.} \textbf{\bibinfo{volume}{125}},
  \bibinfo{pages}{133502} (\bibinfo{year}{2006}).

\bibitem[{\citenamefont{Predescu et~al.}(2005)\citenamefont{Predescu,
  Frantsuzov, and Mandelshtam}}]{Predescu2005}
\bibinfo{author}{\bibfnamefont{C.}~\bibnamefont{Predescu}},
  \bibinfo{author}{\bibfnamefont{P.~A.} \bibnamefont{Frantsuzov}},
  \bibnamefont{and} \bibinfo{author}{\bibfnamefont{V.~A.}
  \bibnamefont{Mandelshtam}}, \bibinfo{journal}{J. Chem. Phys.}
  \textbf{\bibinfo{volume}{122}}, \bibinfo{pages}{154305}
  (\bibinfo{year}{2005}).

\bibitem[{\citenamefont{Miller et~al.}(2003)\citenamefont{Miller, Zhao, Ceotto,
  and Yang}}]{miller2003}
\bibinfo{author}{\bibfnamefont{W.~H.} \bibnamefont{Miller}},
  \bibinfo{author}{\bibfnamefont{Y.}~\bibnamefont{Zhao}},
  \bibinfo{author}{\bibfnamefont{M.}~\bibnamefont{Ceotto}}, \bibnamefont{and}
  \bibinfo{author}{\bibfnamefont{S.}~\bibnamefont{Yang}}, \bibinfo{journal}{J.
  Chem. Phys.} \textbf{\bibinfo{volume}{119}}, \bibinfo{pages}{1329}
  (\bibinfo{year}{2003}).

\bibitem[{\citenamefont{Press et~al.}(1992)\citenamefont{Press, Flannery,
  Teukolsky, and Vetterling}}]{NRC}
\bibinfo{author}{\bibfnamefont{W.~H.} \bibnamefont{Press}},
  \bibinfo{author}{\bibfnamefont{B.~P.} \bibnamefont{Flannery}},
  \bibinfo{author}{\bibfnamefont{S.~A.} \bibnamefont{Teukolsky}},
  \bibnamefont{and} \bibinfo{author}{\bibfnamefont{W.~T.}
  \bibnamefont{Vetterling}}, \emph{\bibinfo{title}{Numerical Recipes in C. The
  Art of Scientific Computing}} (\bibinfo{publisher}{Cambridge University
  Press}, \bibinfo{address}{Cambridge}, \bibinfo{year}{1992}).

\bibitem[{\citenamefont{Liu and Miller}(2006)}]{Liu2006}
\bibinfo{author}{\bibfnamefont{J.}~\bibnamefont{Liu}} \bibnamefont{and}
  \bibinfo{author}{\bibfnamefont{W.~H.} \bibnamefont{Miller}},
  \bibinfo{journal}{J. Chem. Phys.} \textbf{\bibinfo{volume}{125}},
  \bibinfo{pages}{224104} (\bibinfo{year}{2006}).

\bibitem[{\citenamefont{Colbert and Miller}(1992)}]{Colbert91}
\bibinfo{author}{\bibfnamefont{D.~T.} \bibnamefont{Colbert}} \bibnamefont{and}
  \bibinfo{author}{\bibfnamefont{W.~H.} \bibnamefont{Miller}},
  \bibinfo{journal}{J. Chem. Phys.} \textbf{\bibinfo{volume}{96}},
  \bibinfo{pages}{1982} (\bibinfo{year}{1992}).

\bibitem[{\citenamefont{Miller}(1975)}]{MillerTST}
\bibinfo{author}{\bibfnamefont{W.~H.} \bibnamefont{Miller}},
  \bibinfo{journal}{J. Chem. Phys.} \textbf{\bibinfo{volume}{62}},
  \bibinfo{pages}{1899} (\bibinfo{year}{1975}).

\end{thebibliography}
\end{document}